\documentclass{aa}  

\usepackage{graphicx}
\usepackage{txfonts}
\begin{document}

   \title{The dwarf galaxy NGC~185: the flagship of Andromeda's dwarf galaxies for the study of evolved multiple stellar populations}
   \author{P. Ventura\inst{1}, C. Ventura\inst{1}, R. D'Souza\inst{2}, F. Dell'Agli\inst{1}, M. Correnti\inst{1,3}, 
   C. Gavetti\inst{1,4}, \\
   M. Tailo\inst{5}, M. Di Stefano\inst{6}, E. F. Bell\inst{7}}

   \institute{INAF, Observatory of Rome, Via Frascati 33, 00078, Monte Porzio Catone (RM), Italy \\ email: paolo.ventura@inaf.it \and 
   Vatican Observatory, 00120 Vatican City State \and
   ASI-Space Science Data Center, Via del Politecnico, I-00133, Rome, Italy \and
   Dipartimento di Matematica e Fisica, Università degli Studi Roma Tre, 
   via della Vasca Navale 84, 00100, Roma, Italy \and
   Dipartimento di Fisica e Astronomia Augusto Righi, Università degli Studi di Bologna, Via Gobetti 93/2, 40129 Bologna, Italy \and
   Dipartimento di Fisica, Sapienza Università di Roma, Piazzale Aldo Moro 5, I-00185 Roma, Italy \and
   Department of Astronomy, University of Michigan, 1085 S. University Ave, Ann Arbor, MI 48109-1107, USA
   }

   \date{2026}

% \abstract{}{}{}{}{} 
% 5 {} token are mandatory
 
  \abstract
  % context heading (optional)
  % {} leave it empty if necessary  
   {}
  % aims heading (mandatory)
   {Reaching galaxies at 10--15 Mpc with HST and JWST requires star 
   formation history (SFH) diagnostics that do not rely on the faint 
   main sequence turn-off. We investigate whether horizontal branch (HB), 
   red clump (RC), and red and asymptotic giant branch (RGB, AGB) stars can 
   fill this role for dwarf elliptical galaxies, using NGC 185 as a test case. 
   Because the colour and magnitude of HB and RC stars depend on the mass 
   their progenitors lose along the RGB, constraining this mass loss is an 
   essential prerequisite for any SFH derived from these evolved populations.}
  % methods heading (mandatory)
   {We use stellar evolution models of low- and intermediate-mass stars as inputs 
   to population synthesis simulations that reproduce the colour distribution and 
   luminosity function of stars in the $(m_{\rm F606W}-m_{\rm F814W},\, m_{\rm F814W})$ 
   diagram. The study of the stars populating the HB is used to derive 
   the distance modulus and reddening, the average mass loss suffered by the stars during 
   the ascending on the RGB and the SFH occurred during the ancient epochs, before 7 Gyr ago.
   The luminosity function of the RC gives us constraints on the recent SFH ($<8 \mathrm{Gyr}$).
   We then analyse the AGB population,  the sole contributor to the galaxy's dust budget, 
   and combine it with IRAC-Spitzer photometry to characterise the dust properties of objects 
   on the red side of the colour-magnitude diagram.}
  % results heading (mandatory)
   {The distribution of HB stars yields a reddening $E(B-V)=0.20$ mag and a distance 
   modulus $DM=24.10$ mag for NGC 185. The average mass lost by low-mass stars during 
   the RGB phase is $\rm (0.22\pm 0.01)\,M_\odot$. The $5\%$ uncertainty is based on the
   assumption of a single metallicity $\rm Z=0.002$, while possible degeneracy
   including age and helium content would rise the uncertainty to about $10\%$. Star formation 
   was concentrated at early epochs, with half of the stars formed more than 11 Gyr ago. 
   The morphology of the HB indicates essentially no metallicity spread among stars older 
   than 8 Gyr, whereas the slope of the RGB suggests that part of the stars younger than 
   5--6 Gyr are slightly enriched in metals. We compare the resulting SFH with earlier 
   reconstructions based on deep HST photometry reaching the MSTO, and use the modelled 
   AGB population to test the $N_{\rm AGB}/N_{\rm RGB}$--T90 relation proposed for dwarf 
   galaxies. Once RGB mass loss is adequately constrained, evolved stars can provide a 
   reliable SFH for dwarf elliptical galaxies well beyond the reach of MSTO-based methods. 
   We discuss the implications for more distant systems and the remaining uncertainties in 
   the underlying stellar models.}
   % conclusions heading (optional), leave it empty if necessary 
   {}
   

   \titlerunning{Evolved stellar populations of NGC~185}
   \authorrunning{P. Ventura et al.}
   \maketitle
%
%-------------------------------------------------------------------

\section{Introduction}
Significant effort has recently been dedicated to reconstructing the star
formation history (SFH) of individual nearby galaxies through their resolved
stellar populations. Detailed SFHs in the solar neighborhood
\citep{Gallart2024}, Andromeda \citep{Williams2015}, nearby dwarf galaxies
\citep{savino23}, and the stellar halos of nearby MW-mass galaxies
\citep{Brown2006, Rejkuba2005, Durrell2010} have begun to constrain the
physical processes that form and shape galaxies, including gas accretion and
consumption \citep{Cohen2025}, stellar feedback \citep{Hunter2022}, and the
mechanisms that quench star formation, such as the infall of low-mass dwarf
galaxies into the MW \citep[e.g.,][]{Fillingham2015, Slater2014} and cosmic
reionization \citep[e.g.,][]{Benitez-Llambay, Weisz2014b}.

The traditional method for reconstructing galaxy SFHs is fitting the main
sequence turn-off (MSTO) in the colour-magnitude diagram (CMD)
\citep{Brown2006, barker2011, bernard12, bernard15a, bernard15b, monachesi12,
geha15, monelli2016, savino23}. However, the faintness of the MSTO restricts
this method to nearby galaxies ($<$1 Mpc). To reach larger distances
(10--15 Mpc) with HST and JWST, it is important to develop techniques that
derive the SFH from brighter, more evolved populations: stars on the red
giant branch (RGB) and asymptotic giant branch (AGB), and those undergoing
core helium burning on the horizontal branch (HB) and red clump (RC).

Several attempts have used the statistics of evolved stars to constrain
galaxy SFHs. RC stars, together with the RGB stars from which they evolve,
can constrain the SFH over the last 5--6 Gyr \citep{girardi01}: the size and
shape of the RC \citep[e.g.,][]{Caputo1995} and the relative number of RC to
RGB stars \citep[e.g.,][]{Tolstoy1999} both trace the age of the dominant
population. \citet{Martin2017} used HB and RC stars to show that many M31
dwarf galaxies have extended SFHs, inconsistent with rapidly shut-down early
star formation episodes. In recent years, this approach has largely been
superseded by modern CMD-reconstruction methods, which derive the
best-fitting SFH from a galaxy's entire observed CMD, though often with a
systematic mismatch between the observed and modelled RC. \citet{Weisz14}
showed that CMDs reaching only as deep as the HB (including the RC and HB but
not the oldest MSTO) still retain significant SFH information, provided
systematic errors arising from the models are accounted for. The largest
source of uncertainty in age-dating a population via RC stars remains the
uncertainty in mass loss along the RGB ($\rm \delta M_{RGB}$), at the
$\sim20\%$ level.

Along the HB, RR Lyrae periods and amplitudes can also constrain the
building blocks of the Milky Way halo \citep{Fiorentino2015, Fiorentino2017}.
Using the rest of the HB to derive ancient populations has proven more
difficult, limited to a few dwarf galaxies with independent metallicity
constraints. In Carina, \citet{Smecker1994} found two morphologically
distinct horizontal branches, confirming that star formation occurred in two
well-separated episodes at 6 and $>10$ Gyr, later confirmed quantitatively by
\citet{deBoer2014} using MSTO photometry. \citet{salaris13} performed the
first detailed synthetic HB simulation of Sculptor, using an independently
derived SFH and metallicity evolution with RGB mass loss as the only free
parameter, an approach \citet{savino15} extended to Carina, showing HB data
can refine the SFH derived from red giants and turn-off stars alone. This
culminated in the MORGOTH code \citep{savino18}, which showed that including
the HB significantly increases SFH precision; applied to the more distant
Tucana, it resolved three ancient star-formation phases at finer temporal
resolution than previously possible, linked to three discrete HB clumps
\citep{savino19}. While great progress has been made, HB stars work best as
a complement to MSTO/RGB analysis rather than a standalone diagnostic,
remaining fundamentally limited by RGB mass loss — a free parameter current
theory cannot predict from first principles.

A different approach uses the distribution of AGB stars across CMDs to
reconstruct SFHs, enabled by newer AGB models that account for dust
formation in stellar winds and allow reliable SED modelling
\citep{ventura12, ventura14, nanni13, nanni14}. This has been applied to the
Magellanic Clouds \citep{flavia14, flavia15a, flavia15b}, several Local Group
galaxies \citep{flavia16, flavia18, flavia19, cla26}, and Andromeda
\citep{cla25}, since circumstellar dust mineralogy and quantity depend
strongly on the mass and metallicity of the AGB progenitor \citep{ventura14},
linking individual CMD sources to a formation epoch and composition.

\citet{harmsen23} showed that in smaller dwarf galaxies, the number ratio of
AGB to RGB stars ($\rm N_{AGB}/N_{RGB}$), computed in boxes above and below
the tip of the RGB (TRGB) in the $\rm (m_{F606W}-m_{F814W}, m_{F814W})$
diagram, constrains T90 (the time within which $90\%$ of a galaxy's stars
formed), calibrated against dwarf galaxies with SFHs from MSTO stars.
\citet{ventura26} tested this method against RGB/AGB stellar evolution
models, confirming the expected $\rm N_{AGB}/N_{RGB}$--T90 anti-correlation
but showing its slope depends critically on $\rm \delta M_{RGB}$.

For larger galaxies, \citet{Lee2025} showed that a spatially resolved SFH
from $\sim$7,700 AGB stars modelled with the COLIBRI models
\citep{Marigo2017, Pastorelli2019, Pastorelli2020} closely matches deeper
optical results from $\sim$33 million stars in the Panchromatic Hubble
Andromeda Treasury survey \citep{Williams2017}. In smaller galaxies like WLM,
the same technique \citep{Lee2026} qualitatively reproduces the SFH from
deeper JWST data \citep{Cohen2025} reaching the old MSTO, though limited by shot
noise from few AGB stars. Such techniques remain limited by the underlying
models and mass-loss assumptions.
Long-period variable stars, monitored in the K band, offer another route to
degeneracies and shot noise from the short AGB lifespan.

Reconstructing SFHs from AGB populations works best for larger systems with
recent star formation and larger AGB samples. It is less effective for
dwarf elliptical galaxies, where the AGB population is often limited to a
few dozen stars and old, low-mass stars produce fainter AGB stars with
negligible dust, causing CMD tracks to run nearly vertically and overlap in
the optical \citep{cla26}. Nevertheless, modelling the AGB population offers
an additional constraint on RGB mass loss. A more accurate SFH for
lower-mass galaxies requires extending the analysis to the more numerous HB,
RC, Early-AGB (E-AGB), and RGB stars, provided adequate constraints exist on RGB mass
loss.

With improved models and tighter constraints on red giant branch (RGB) mass loss 
now available, we revisit the use of evolved stars, namely horizontal branch (HB) 
and red clump (RC) stars, to reconstruct the star formation histories (SFHs) of 
dwarf elliptical galaxies. The magnitude and colour of HB stars depend primarily 
on metallicity, age, and RGB mass loss. Helium abundance variations associated with 
multiple stellar populations, often invoked in globular cluster studies, have a much 
smaller effect on the HB in these systems and are neglected here. We investigate how 
well the colour and magnitude distributions of HB and RC stars, combined with the 
luminosity functions (LFs) of RGB and AGB stars, can trace the SFHs of these galaxies.

We focus on NGC 185 ($M_V= -15.6$), which hosts a substantial HB and RC population, making
it an ideal test case. Its SFH has previously been reconstructed from deep
HST photometry reaching the MSTO \citep{geha15} and from its long-period
variables \citep{javadi17}. \cite{savino23} used RR Lyrae stars for a
robust distance modulus ($24.06 \pm 0.06$), while \citet{mahani25} studied dust production by
the long-period variables of \citet{javadi17}. Recent star formation has
also been inferred from red supergiant stars found by \citet{ren21}.
NGC 185 is also among the galaxies used by \citet{ventura26} to critically
test the T90--$\rm N_{AGB}/N_{RGB}$ relationship of \citet{harmsen23}.

We used stellar evolution models of low- and intermediate-mass stars as
inputs to population synthesis simulations reproducing the colour
distribution and LF of stars in the $\rm (m_{F606W}-m_{F814W}, m_{F814W})$ 
diagram. AGB stars, though far outnumbered by RGB, HB, and RC stars, 
are crucial both for testing the $\rm N_{AGB}/N_{RGB}$--T90 method 
\citep{harmsen23} and for the galaxy's dust budget, to which they are 
the sole contributors.

The paper is structured as follows: Section~\ref{input} describes the
population synthesis inputs; Section~\ref{evol} discusses the properties of
stars in the post-core-hydrogen-burning phase; Section~\ref{sfh} presents
the population synthesis results and derived SFH; Section~\ref{agb} covers
the galaxy's AGB population and the dust properties of objects on the red
side of the CMD from IRAC-Spitzer photometry; Section~\ref{disc} discusses
the results; the conclusions are given in Section~\ref{concl}.

\section{Physical and numerical input}
\label{input}
The analysis of NGC 185 was carried out in several steps: modelling 
stellar evolution, describing the dust formation process in the stellar 
wind (required in particular for interpreting the Spitzer data), and 
applying a population synthesis approach to generate the synthetic 
distribution of stars across the various colour-magnitude diagrams 
for comparison with the observations. The methodology is described 
in detail in \citet{cla25}; the only difference here is that the SFH 
adopted as the starting point for the interpretation was taken from 
\citet{savino25}.

We use the original ACS data from \citet{geha15}, reprocessed by \citet{savino25}\footnote{Although the original HST GO-11724 proposal (PI: Geha) 
included 30-s exposures, they were not included in the final data reduction stack.}, 
which reach below the oldest MSTO  (assuming a metallicity of $\rm [Fe/H] 
\sim -1$ for NGC 185), albeit with significant colour uncertainties at the 
oldest ages. The ACS field is located 6'.6 from the centre of NGC185 
corresponding to 2.5 effective radii: the stellar density at this radius does not
permit us to detect an appreciable number of AGB stars. The observations also have 
enough temporal cadence in F606W and F818W data to reconstruct the light curves 
of the RR Lyrae stars \citep{monelli2017}.

\subsection{Stellar evolution modelling}
\label{starmod}
Based on the $\rm [Fe/H] \sim -1$ metallicity of NGC 185 stars claimed in previous
studies \citep{geha15}, we based most of the analysis on the stellar models 
of metallicity $\rm Z=2\times 10^{-3}$ taken from the Rome library, published in
\citet{devika23}. To explore the possibility of different chemistries or
of some metallicity spread, we based part of the analysis on stellar models
of metallicity $\rm Z=10^{-3}$ and $\rm Z=4\times 10^{-3}$, also published
in \citet{devika23}.

The stellar models, calculated by means of the ATON code
for stellar evolution \citep{italo, ventura98}, were evolved from the pre-MS
through the core hydrogen burning, RGB, E-AGB phases\footnote{We refer to 
E-AGB as the evolutionary phases running from the exhaustion of central helium until the
ignition of the first thermal pulse.}, then for a few thermal pulses.
For $\rm M<2~M_{\odot}$ stars, the computations were stopped at the ignition
of the helium flash, then resumed from the core helium burning phase. We note
that the mass with which the stars begin the core helium-burning phase is
generally different from the initial mass of the star, being sensitive
to $\rm \delta M_{RGB}$, the amount of mass lost during the ascent of the RGB.
This parameter was shown to play a crucial role in the synthetic modelling
of the evolved stellar population of galaxies \citep{cla25}, and to deeply
affect the estimate of the T90 of the galaxies mainly composed by an old stellar
population \citep{ventura26}.
In the following, we will distinguish between $\rm M$, the initial mass of the
star, and $\rm M_{HB}$, the mass with which the core helium burning phase begins.

\subsection{Dust formation and the evolution of the spectral energy distribution}
The results from stellar evolution modelling, particularly the values
of mass, effective temperature, luminosity, rate of mass loss, and
surface chemical composition, are used as ingredients to model
dust production, by means of the schematization
described in detail in \citet{fg02, fg06}. The wind is assumed to be stationary
and to expand isotropically from the surface of the star, where 
dust formation occurs when the temperatures drop below the stability threshold
at which the rates of vaporization and of the grain growth balance each 
other, typically below $\sim 1500$ K. The dust species considered are
silicates and alumina dust, in oxygen-rich environments, and amorphous carbon
and silicon carbide, in the circumstellar envelope of carbon stars. All the
relevant equations used to simulate the formation and the growth of the dust
grains in the wind are described in \citet{ventura12}. The results from the
modelling of dust formation, particularly the density stratification of the
wind, the mineralogy and the size distribution of the dust formed, and the
optical depth, are the ingredients required by the code DUSTY \citep{nenkova99},
which allows the computation of the SED: this step
is necessary to follow the change in the position of the stars on the 
CMDs built with near- and mid-IR filters.

\subsection{Population synthesis}
\label{sintmod}
The stellar models mentioned earlier in this section are used in a 
population synthesis approach, aimed at producing synthetic 
distributions of NGC 185 stars on the $\rm (m_{F606W}-m_{F814W}, m_{F814W})$ CMD, 
to be compared with the HST data from \citet{savino25}. As a starting point, 
we consider star formation starting from 14 Gyr ago to the present, 
and extract for each epoch considered a number of stars, based on the SFH 
given in \citet{savino25}. We weight the number of stars extracted assuming the 
Kroupa initial mass function \citep{kroupa01}. 
The SFH is then adjusted until reproducing the observed number counts
of NGC 185 stars in the various regions of the CMD.
The number of RGB, core helium burning, E-AGB and AGB stars considered is 
derived on the basis of the evolutionary timescales of the model stars of different
mass and on the duration of the individual phases. The outcome
of this analysis was the characterization of the stellar sources in NGC 185,
the determination of the SFH, the RGB mass loss $\rm \delta M_{RGB}$,
reddening and distance modulus that allow the best reproduction of the
data.

\section{The evolution of the stars across the post core hydrogen
burning phases}
\label{evol}

\begin{figure}
\begin{minipage}{0.48\textwidth}
\resizebox{1.\hsize}{!}{\includegraphics{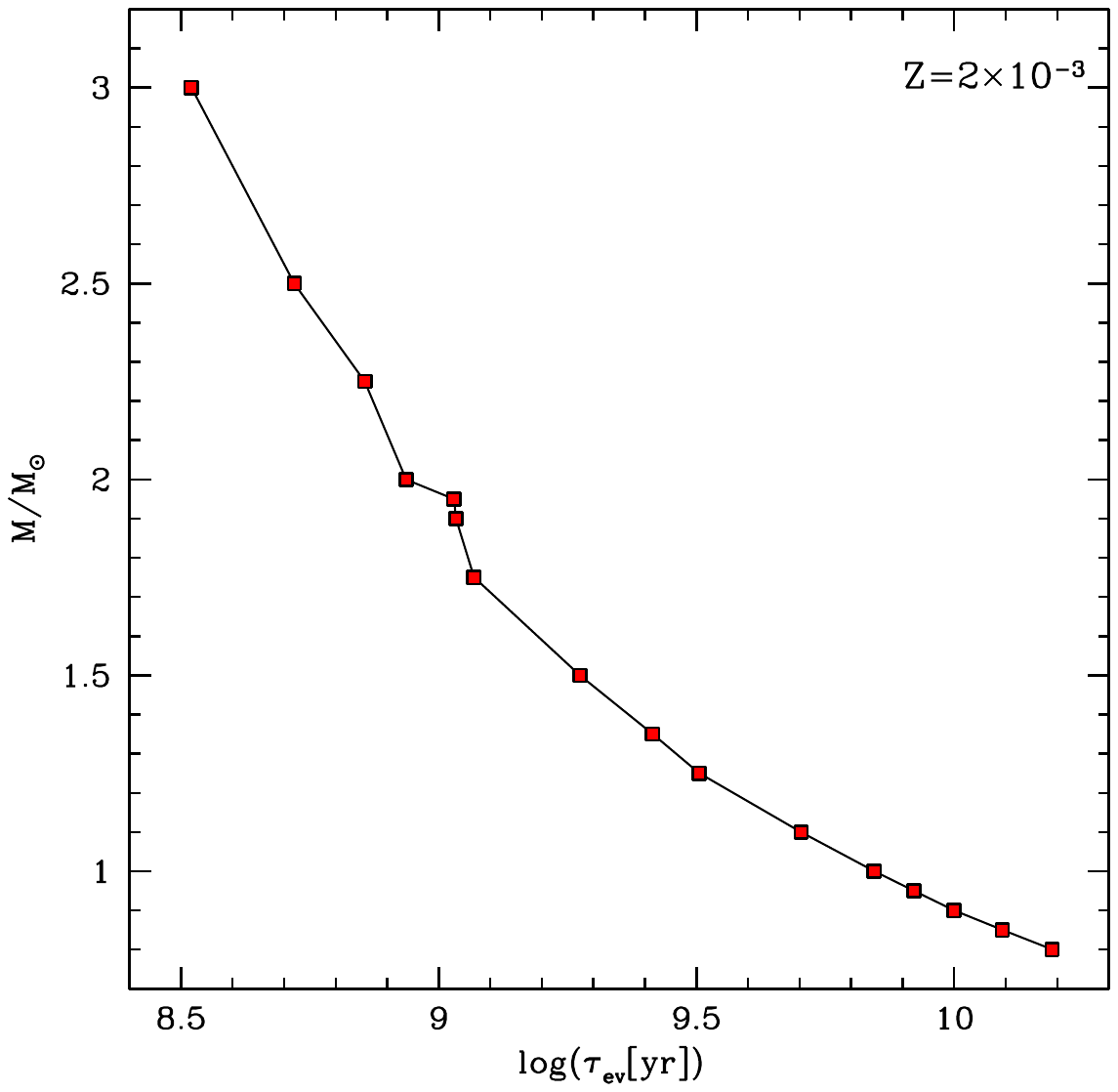}}
\end{minipage}
\vskip-60pt
\caption{Evolutionary time scales of model stars of metallicity
$\rm Z=2\times 10^{-3}$ and mass in the $\rm 0.8-3~M_{\odot}$
range.
}
\label{ftempi}
\end{figure}

The evolved population of galaxies is formed by stars born in different
epochs, nowadays evolving through the different post core hydrogen burning 
phases. Since the evolutionary time scales of the stars decrease with
increasing mass, the current population of the galaxy is a mixture of
old, low-mass stars and younger, higher mass objects. Fig.~\ref{ftempi} 
shows the evolutionary time-scales $\rm \tau_{ev}$ of $\rm Z=2\times 10^{-3}$ 
stars of different mass. We do not consider $\rm M < 0.8~M_{\odot}$ stars,
because they evolve on time scales longer than 14 Gyr. We also exclude the 
stars of mass above $\rm 3~M_{\odot}$, since they evolve on time scales  
shorter than 300 Myr, while the present analysis can be applied
with significant statistical robustness only to study
epochs older than $\sim 800$ Myr.
It is clear in Fig.~\ref{ftempi} the negative trend 
of $\rm \tau_{ev}$ with mass: $\rm \tau_{ev}$ descreases from 
about 10 Gyr to 3 Gyr and 1 Gyr, as the stellar mass increases from $\rm 0.9~M_{\odot}$, 
to $\rm 1.25~M_{\odot}$ and $\rm 2~M_{\odot}$.

\begin{figure}
\begin{minipage}{0.48\textwidth}
\resizebox{1.\hsize}{!}{\includegraphics{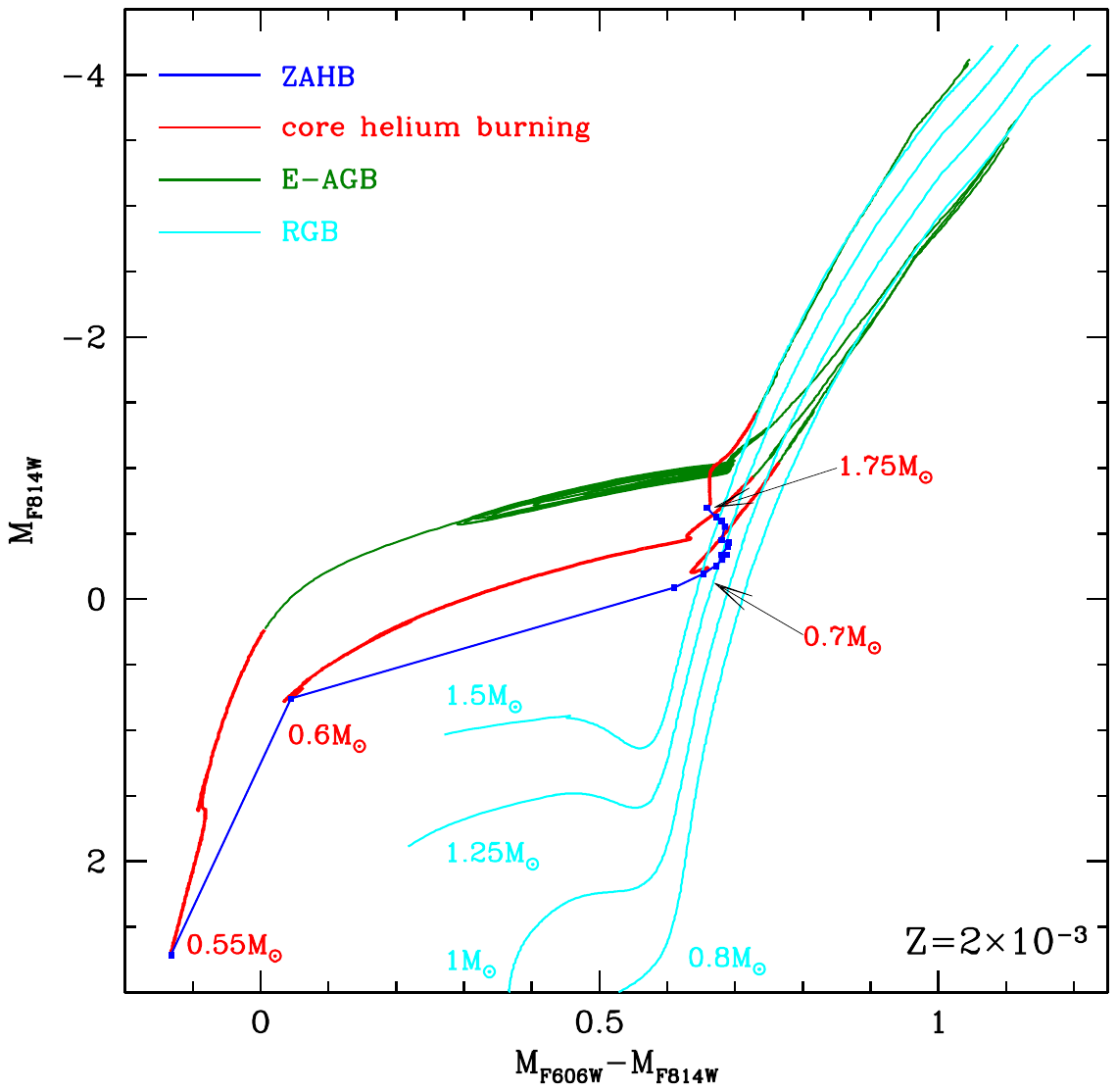}}
\end{minipage}
\vskip-60pt
\caption{Evolutionary tracks of $\rm Z=2\times 10^{-3}$ model stars
of different mass, on the $\rm (m_{F606W}-m
_{F814W}, m_{F814W})$
diagram. Cyan tracks refer to the post core H-burning evolution of 
stars of initial mass $\rm 0.8, 1, 1.25, 1.5~M_{\odot}$ (indicated 
in the figure), extended until the ignition of the helium flash. 
In the figure we also report the tracks corresponding to the
evolution started from the core helium burning phase of stars
of mass (taken at the start of the core helium burning)
$\rm 0.55, 0.6, 0.7, 1.75~M_{\odot}$, extended until the start
of the AGB phase. The tracks are coloured in red until central
helium is exhausted, then the dark green parts refer to the
E-AGB phase. The blue line indicates the Zero Age Horizontal Branch,
connecting the points representing the start of the core helium
burning phase of model stars of mass in the $\rm 0.55-1.75~M_{\odot}$
range.
}
\label{ftracce}
\end{figure}

Fig.~\ref{ftracce} shows the evolutionary tracks of stars of
different mass in the $\rm (m_{F606W}-m_{F814W}, m_{F814W})$
diagram. The cyan lines refer to the RGB phase of model stars
of different mass, and are extended until the TRGB, which, for 
the $\rm Z=2\times 10^{-3}$ metallicity 
discussed here, is found at $\rm M_{F814W} \sim -4.21$ mag. The tracks 
develop towards the red side of the plane, until reaching
colours $\rm (m_{F606W}-m_{F814W}) \sim 1.1$ mag, with 
a $\sim 0.2$ mag spread, related to the stellar mass, with
higher mass stars being bluer. The highest mass of the stars
undergoing the helium flash for the metallicity discussed
here is $\rm 1.9~M_{\odot}$. The duration
of the RGB evolution ($\rm \tau_{RGB}$), particularly of the 
evolutionary phases during which the tracks develop almost 
vertically on the CMD, is sensitive to the initial mass of the star:
$\rm \tau_{RGB} \sim 1.3 $ Gyr for $\rm M=0.8~M_{\odot}$,
$\rm \tau_{RGB} \sim 800 $ Myr for $\rm M=1~M_{\odot}$,
$\rm \tau_{RGB} \sim 70 $ Myr for $\rm M=1.9~M_{\odot}$.
Stars with $\rm M > 1.90~M_{\odot}$ undergo quiescent helium burning:
the duration of the RGB phase until the ignition of helium in
the core is shorter than 20 Myr (with some dependence on the
initial mass), thus they are not expected to contribute
significantly to the RGB population of the galaxy.

In Fig.~\ref{ftracce} we also report the evolutionary tracks 
of a few model stars, for which the computations were started
from the beginning of the quiescent core helium burning phase.
The part of the tracks coloured in red indicates the evolutionary
phases until the exhaustion of central helium, whereas the
dark green line corresponds to the E-AGB evolution. The blue line
in the diagram indicates the Zero Age Horizontal Branch (ZAHB), 
connecting the position of the
model stars of different mass that have just started core helium
burning. 

We note the large sensitivity of the 
position of the stars undergoing core helium burning on the CMD to $\rm M_{HB}$.
This is particularly clear in the $\rm 0.55~M_{\odot} < M_{HB} < 0.7~M_{\odot}$
domain, where a gap of $\rm \delta (m_{F606W}-m_{F814W}) \sim 0.7$ mag 
separates the position of the $\rm M_{HB}=0.55~M_{\odot}$ and of
the $\rm M_{HB}=0.7~M_{\odot}$
model stars. We also note the large $\rm m_{F814W}$ difference, 
exceeding 2.5 mag, between the two model stars, due to the fact that
the $\rm M_{HB} = 0.55~M_{\odot}$ model star is almost entirely supported by core
helium burning, with negligible contribution from the CNO nuclear
activity in the shell.
The impact of $\rm M_{HB}$ on the position along the ZAHB is notably reduced 
in the $\rm M_{HB} \geq 0.7~M_{\odot}$ domain, where the different model 
stars tend to cluster around $\rm (m_{F606W}-m_{F814W}) \sim 0.7$ mag,
$\rm m_{F814W} \sim -2.5$ mag, which favours the formation of a
clump on the CMD.

The duration of the core helium burning phase ($\rm \tau_{HB}$) is slightly 
shorter than 100 Myr in the sub-solar mass domain, almost independently of
$\rm M_{HB}$. On the other hand, $\rm \tau_{HB}$ is found to decrease for 
model stars of mass above solar, from $\rm \tau_{HB} = 9\times 10^7$ yr, 
for $\rm M_{HB}=1~M_{\odot}$, to $\rm \tau_{HB} = 8.4\times 10^7$ yr, for 
$\rm M_{HB}=1.9~M_{\odot}$.

\begin{figure}
\begin{minipage}{0.48\textwidth}
\resizebox{1.\hsize}{!}{\includegraphics{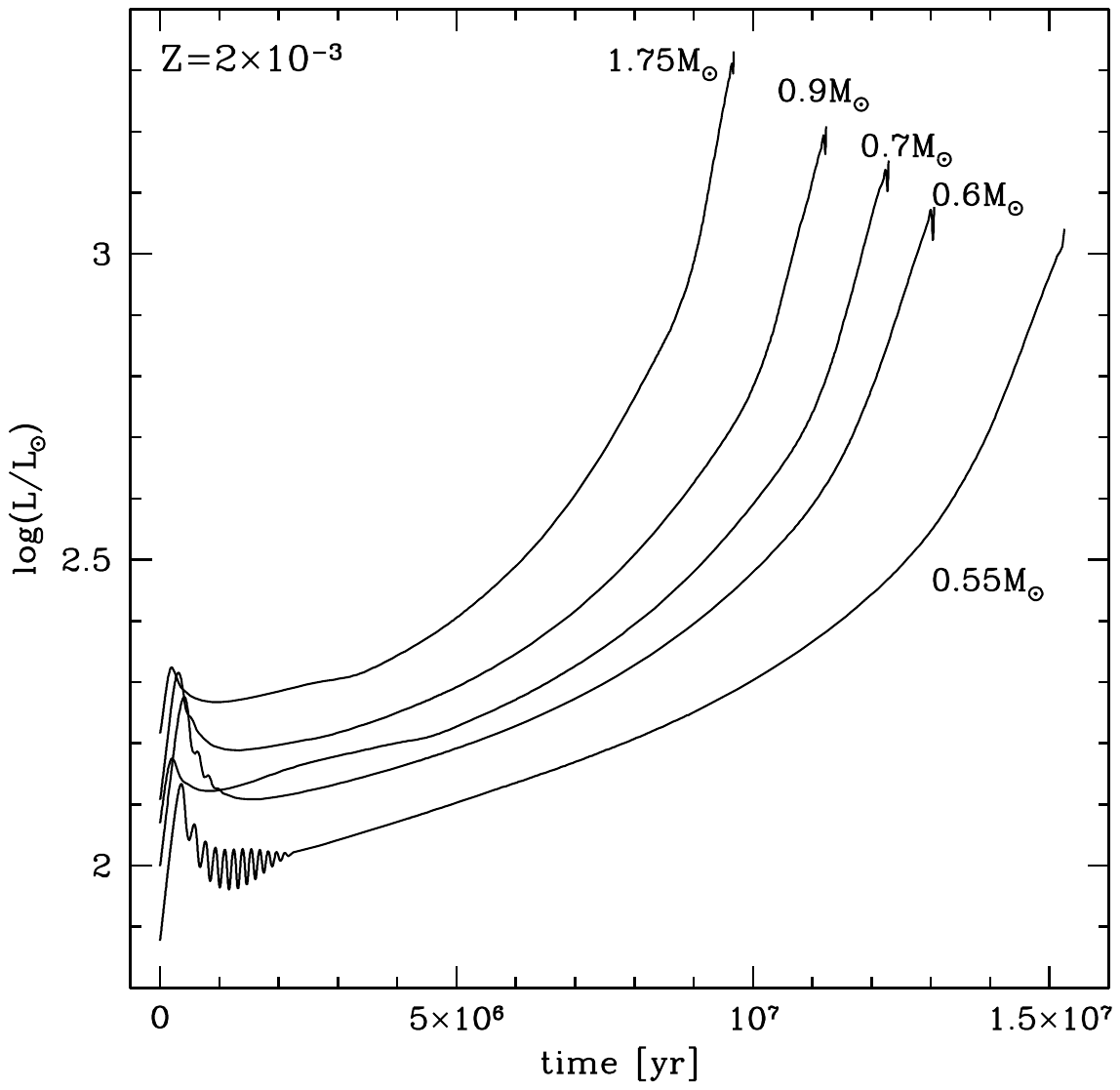}}
\end{minipage}
\vskip-60pt
\caption{Time evolution of the luminosity of model stars
of mass $\rm M_{HB}$ in the $\rm 0.55-1.75~M_{\odot}$ domain,
during the E-AGB phase. The tracks start in conjunction
of the exhaustion of central helium, and are terminated 
shortly before the ignition of the first thermal pulse.
Times are zeroed at the exhaustion of central helium.
}
\label{feagb}
\end{figure}

Fig.~\ref{feagb} shows the time evolution of the luminosity
of model stars of different mass, limited to the E-AGB phase.
The sequences start when the central helium is exhausted, and
are interrupted at the occurrence of the first thermal pulse. 
As a consequence of the larger core mass attained at the end
of core helium burning, higher mass models evolve faster,
so that the duration of the E-AGB phase decreases from 
$\sim 1.5\times 10^7$ yr, for $\rm M_{HB}=0.55~M_{\odot}$, to
slightly shorter than $10^7$ yr, for $\rm M_{HB}=1.75~M_{\odot}$.

During the initial part of the E-AGB the luminosity undergoes a
relatively fast increase, then decreases, when the CNO activity in
the shell is temporarily extinguished. Restore of the H-burning
activity triggers a further increase in the luminosity, which
eventually reaches $\rm L \sim 1-2 \times 10^3~L_{\odot}$, before
the start of the AGB. The results reported in Fig.~\ref{feagb} 
indicate that the stars evolving through the E-AGB spend a significant
fraction of their time during the phase when the luminosity is
at the above mentioned local minimum occurring at the extinction
of the H-burning shell: we therefore expect that most of the E-AGB 
stars are now evolving at luminosities within the 
$\rm 1-2 \times 10^2~L_{\odot}$ range, populating a region of
the CMD $2-3$ mag fainter than the TRGB.

\begin{figure*}
\begin{minipage}{0.44\textwidth}
\resizebox{1.\hsize}{!}{\includegraphics{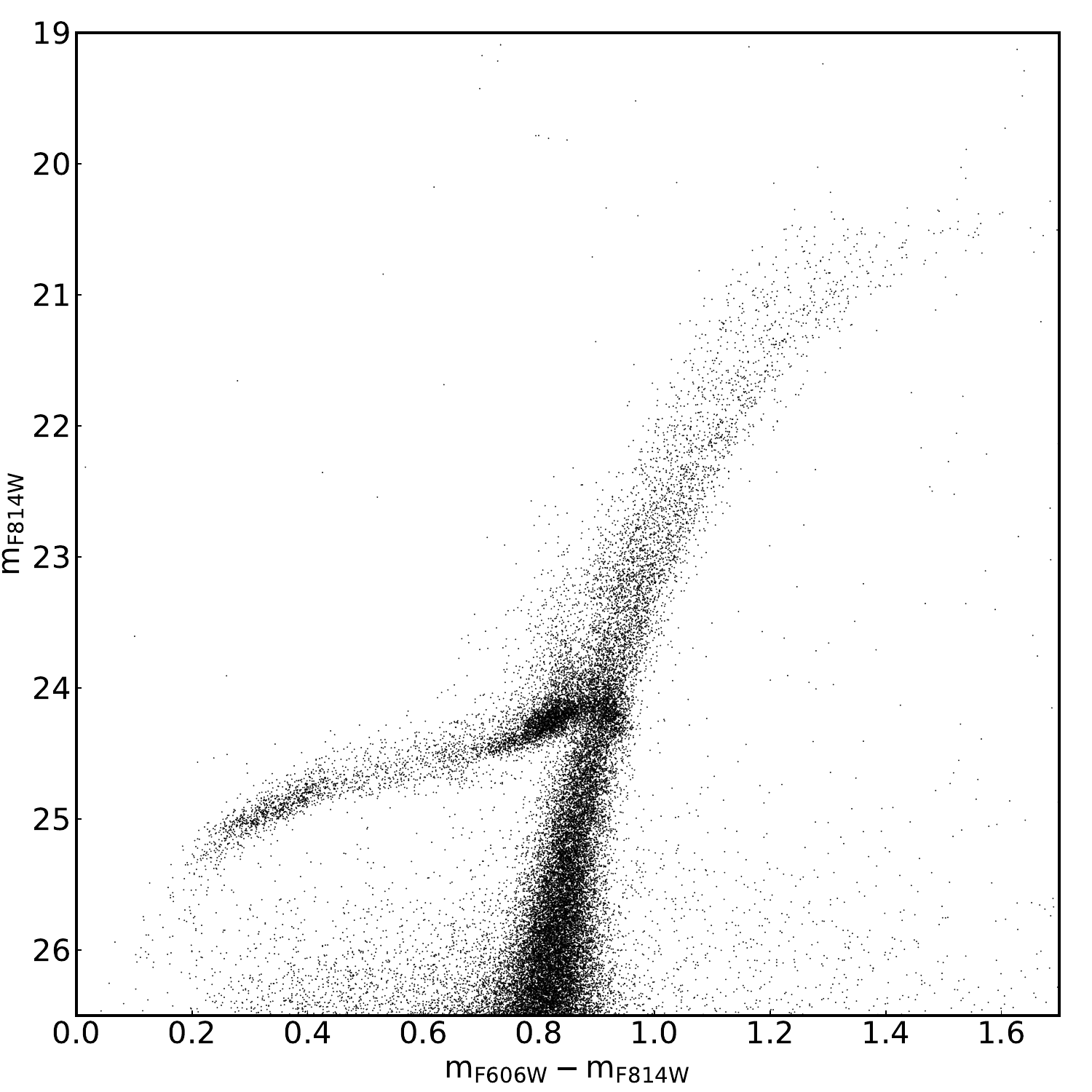}}
\end{minipage}
\begin{minipage}{0.44\textwidth}
\resizebox{1.\hsize}{!}{\includegraphics{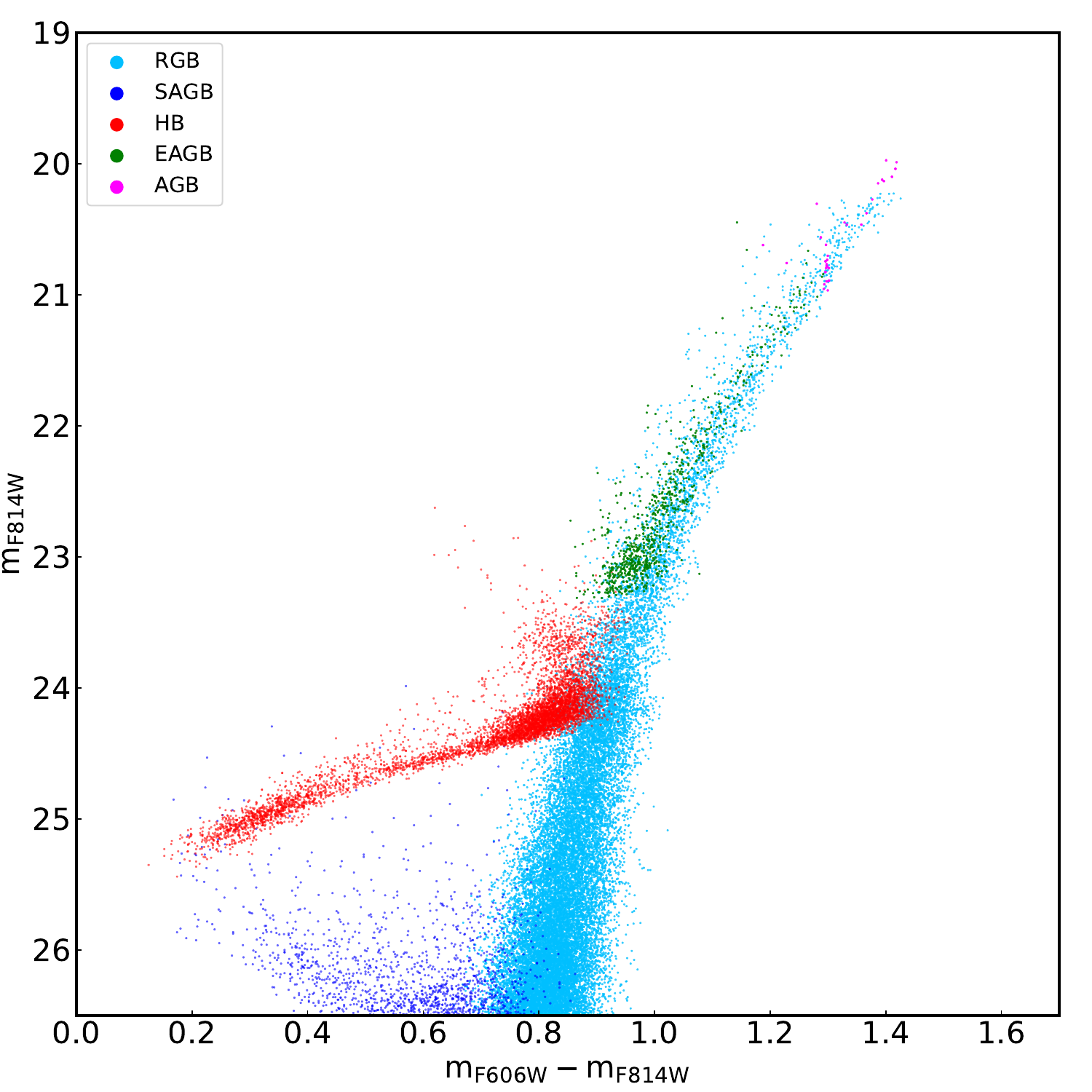}}
\end{minipage}
\vskip-0pt
\caption{Left: Evolved stellar population belonging to the galaxy NGC~185
on the $\rm (m_{F606W}-m_{F814W}, m_{F814W})$ diagram. Right:
Distribution of NGC~185 evolved stars obtained by the 
population synthesis method described in section \ref{sintmod}. 
Stars evolving through the core helium burning, RGB, SGB, E-AGB 
and AGB 
phases are distinguished on the basis of the colour coding
reported in the legend.
} 
\label{fcmd}
\end{figure*}

\section{Reconstructing the star formation history of NGC~185}
\label{sfh}
The way we followed to reconstruct the SFH
of NGC 185 runs through the interpretation of the distribution
of the evolved stars of the galaxy on the $\rm (m_{F606W}-m_{F814W}, m_{F814W})$ CMD, 
shown in the left panel of Fig.~\ref{fcmd}. We note three distinct features,
namely: a) the RGB sequence, extending until the TRGB, at 
$\rm m_{F814W} \sim 20.3$ mag; b) HB stars, distributed in the
colour range $\rm 0.2~\rm{mag} < (m_{F606W}-m_{F814W}) < 0.7~\rm{mag}$; 
c) the RC of core helium burning stars, distributed along the
strip located at $\rm (m_{F606W}-m_{F814W}) \sim 0.84 ~\rm{mag}$,
extended across the range $\rm 23~\rm{mag} < m_{F814W} < 24.5$ mag.

The outcome of this analysis is the synthetic distribution
shown in the right panel of Fig.~\ref{fcmd}, where the individual
sources are colour-coded according to their evolutionary status:
core helium burning objects are shown in red, while RGB, E-AGB, AGB 
and sub-giant branch (SGB) stars are indicated with cyan, dark green, 
magenta and blue points, respectively.

\subsection{Reading the ancient star formation of the galaxy through the HB}
\begin{figure}
\begin{minipage}{0.48\textwidth}
\resizebox{1.\hsize}{!}{\includegraphics{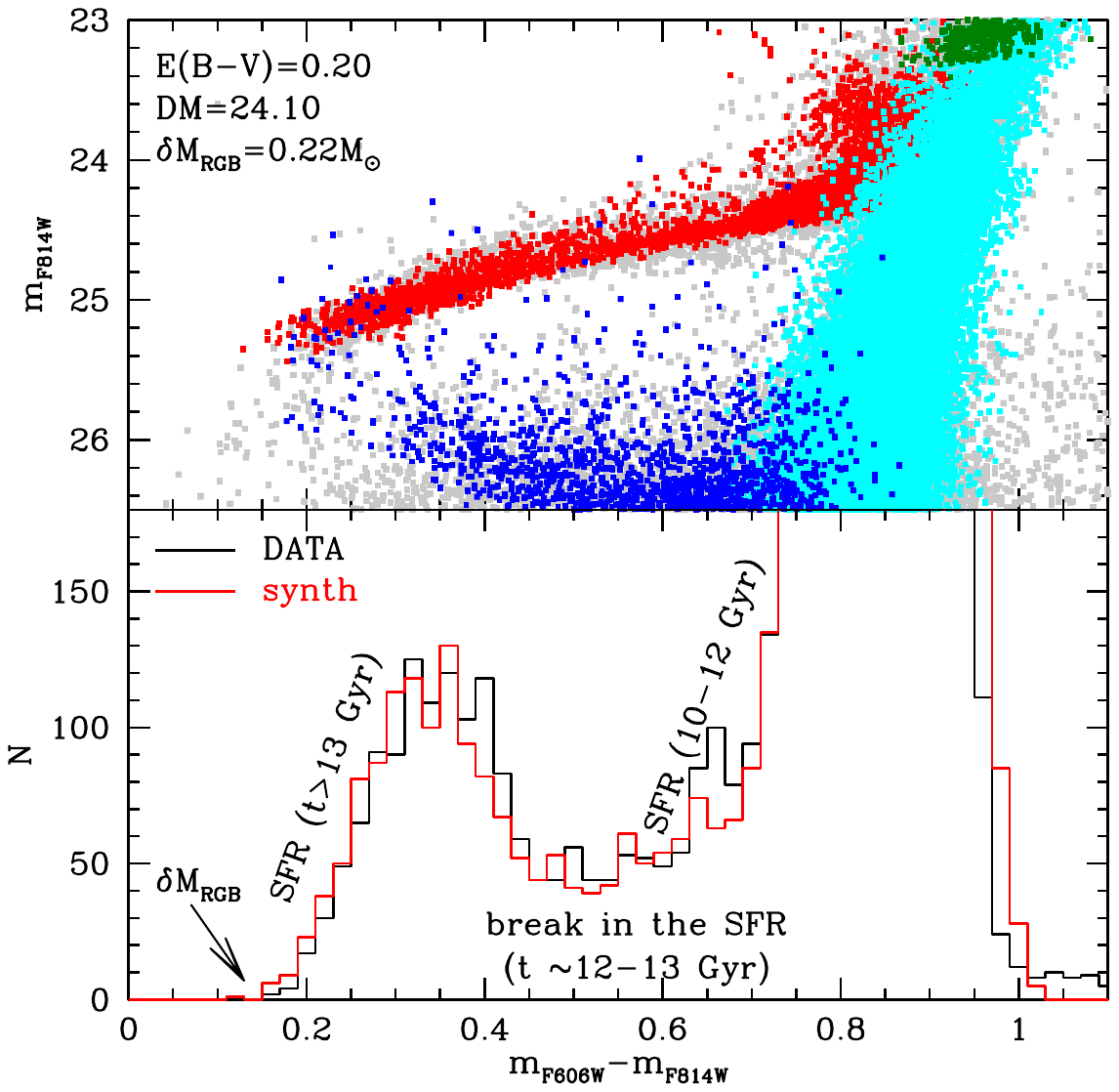}}
\end{minipage}
\vskip-60pt
\caption{Top: The sinthetic distribution of NGC~185 stars in the
region of the HB overlayed to the HST data points, shown in grey.
The colour coding of the synthetic stars is the same as in the
right panel of Fig.~\ref{fcmd}.
Bottom: comparison between the observed colour distribution of 
NGC~185 stars populating the region of the 
$\rm (m_{F606W}-m_{F814W}, m_{F814W})$ diagram within the
$\rm 24~\rm{mag} < m_{F814W} < 25.5~\rm{mag}$ strip and the
synthetic distribution. 
}
\label{fhb}
\end{figure}

The starting point of the present analysis is the interpretation
of the sources populating the HB of the galaxy, corresponding to
the point b) listed above. A zoom of the region of the CMD populated
by HB stars is shown in the top panel of Fig.~\ref{fhb}. 
As discussed in section \ref{evol}
(see the blue line in Fig.~\ref{ftracce}), this region 
of the CMD is populated by stars in the core helium burning
phase, with mass $\rm M_{HB}$ in the $\rm 0.55-0.7~M_{\odot}$ range.
These objects descend from the stars of lowest mass among the
evolved stellar population of the galaxy, and
can be used to reconstruct the star formation that 
occurred in the most ancient epochs, before $\sim 10$ Gyr ago. 

The notable sensitivity of the position of HB stars
on the $\rm (m_{F606W}-m_{F814W}, m_{F814W})$ 
CMD to the stellar mass, clear from the morphology of the ZAHB
reported in Fig.~\ref{ftracce}, renders the analysis of
the colour and magnitude distribution of HB stars the
most valuable indicator not only of the SFH experienced 
during the remote times, but also of general properties 
of the galaxy. Indeed the simultaneous fit of the stars distribution 
populating the whole HB and the RGB region in the
$\rm 24~\rm{mag} < m_{F814W} < 25.5~\rm{mag}$ strip can be 
obtained with an almost unique combination of the reddening 
and distance modulus of the galaxy, of the details of the SFH 
occurred in ages older than 8 Gyr, and of the average mass loss 
$\rm \delta M_{RGB}$ experienced by the stars while ascending the RGB.

The technique adopted can be understood by inspection of  
the bottom panel of Fig.~\ref{fhb}, where the $\rm (m_{F606W}-m_{F814W})$ colour 
distribution of NGC 185 stars in the 
$\rm 24~\rm{mag} < m_{F814W} < 25.5~\rm{mag}$
region of the diagram is compared with the best
fit that we obtained by synthetic modelling. The blue limit of the
distribution, located at $\rm (m_{F606W}-m_{F814W}) \sim 0.15~\rm{mag}$,
is consistent with an average RGB mass loss $\rm \delta M_{RGB}=0.22~M_{\odot}$.
The simultaneous fit of the numerical consistency of the two most populated 
branches of the HB and of the colour distribution of RGB stars 
leads to the estimate of the distance modulus $\rm DM=24.10$ mag and reddening 
$\rm E(B-V)=0.20$ mag. The height of the blue peak in the colour
distribution, visible in the bottom panel of Fig.~\ref{fhb} 
at $\rm (m_{F606W}-m_{F814W}) \sim 0.35~\rm{mag}$, 
provides an estimate of the star formation rate (SFR) 
in the epochs older than 13 Gyr, which was 
$\rm \sim 4\times 10^{-3}~M_{\odot}/$yr, while the minimum centered at 
$\rm (m_{F606W}-m_{F814W}) \sim 0.5~\rm{mag}$ indicates a stop in the star 
formation activity between 12 and 13 Gyr ago. 
Finally, the slope of the colour distribution in the 
$\rm 0.5~\rm{mag} < (m_{F606W}-m_{F814W}) < 0.7~\rm{mag}$ range
is indicative of the efficiency of the star formation process 
between 8 and 12 Gyr ago, which on the average was
$\rm \sim  10^{-3}~M_{\odot}/$yr. The distribution of the stars
across the HB region is so sensitive to the SFH occurred
until $\sim 8$ Gyr ago, that the latter can be reconstructed
with an accuracy within $10\%$.

\begin{figure*}
\vskip-40pt
\begin{minipage}{0.44\textwidth}
\resizebox{1.\hsize}{!}{\includegraphics{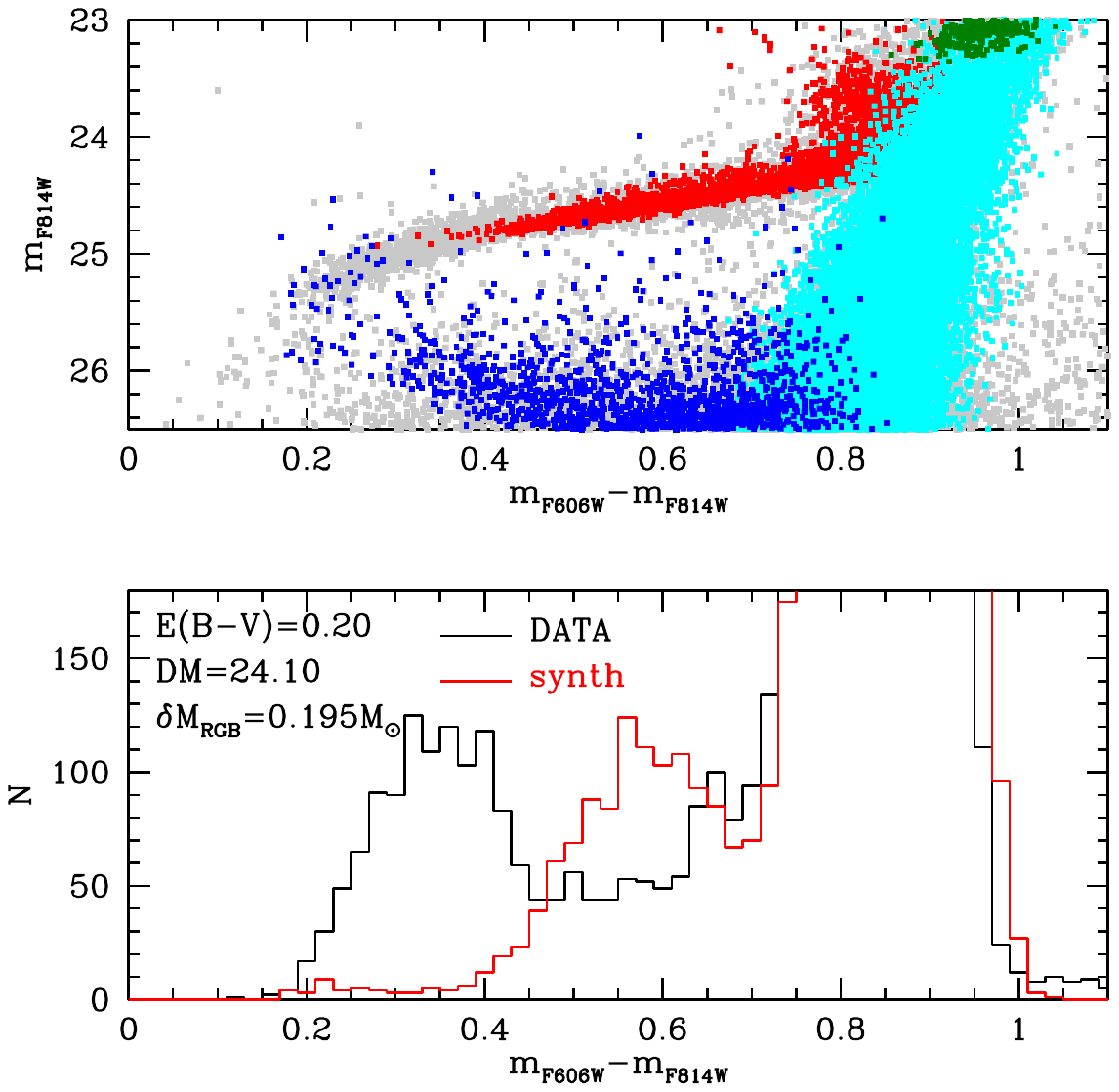}}
\end{minipage}
\begin{minipage}{0.44\textwidth}
\resizebox{1.\hsize}{!}{\includegraphics{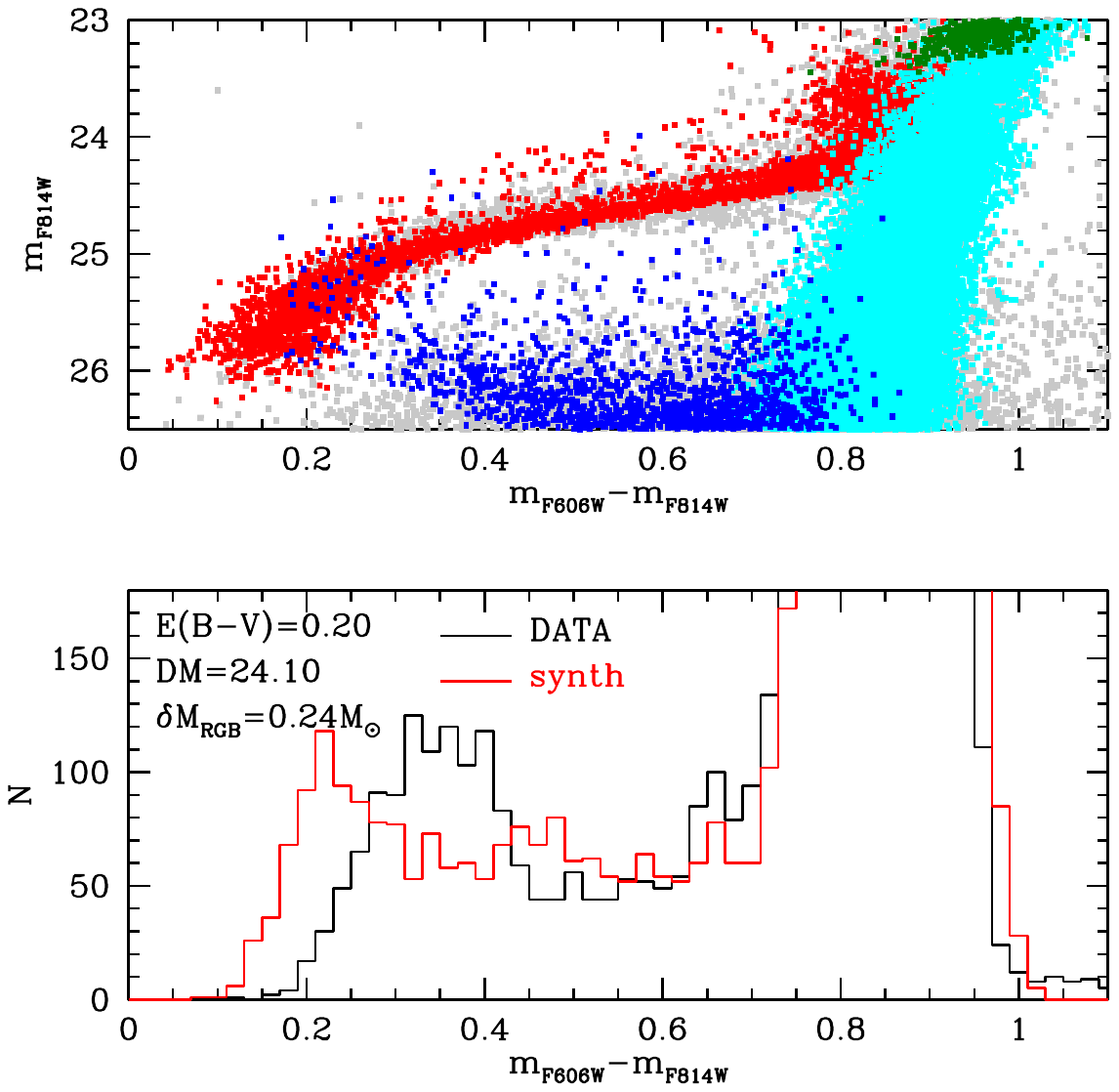}}
\end{minipage}
\vskip-60pt
\caption{Comparison between the observed and the synthetic 
distributions of HB stars belonging to NGC~185 across the colour-magnitude
$\rm (m_{F606W}-m_{F814W}, m_{F814W})$ diagram (upper panels),
and of the corresponding colour histograms (lower panels),
when RGB mass losses of $\rm \delta M_{RGB}=0.195~M_{\odot}$ (left)
and $\rm \delta M_{RGB}=0.24~M_{\odot}$ (right) are adopted.
} 
\label{fdmrgb}
\end{figure*} 

To show the high sensibility of these results on the assumed
RGB mass loss, we report in Fig.~\ref{fdmrgb} the
results obtained by assuming $\rm \delta M_{RGB}$ 
$10\%$ different from the $\rm 0.22~M_{\odot}$ value 
given above. When a $10\%$ lower value of 
$\rm \delta M_{RGB}$ is adopted, the sequence of core
helium burning stars extends 
to $\rm (m_{F606W}-m_{F814W}) \sim 0.4$ (top, left panel), 
thus failing to explain the bluest part of the HB. 
On the other hand, when a $10\%$ higher RGB mass loss 
is assumed (top, right panel), the HB extends to 
$\rm (m_{F606W}-m_{F814W}) \sim 0.05$, with a
significant number of blue stars in the $\rm m_{F814W} \sim 26$
region of the plane. The bottom panels of Fig.~\ref{fdmrgb}
shows the comparison between the observed and the synthetic 
colour distributions of the HB stars, obtained in the
two cases discussed. The results outline severe inconsistencies 
with the observational evidence in either cases, thus the 
assumed $\rm \delta M_{RGB} = 0.22~M_{\odot}$ can be barely changed
by more than $5\%$. This is a lower limit to the uncertainty
on the estimate of $\rm \delta M_{RGB}$, as further effects 
related to the combination between age, a possible metallicity 
spread and the helium content might lift the overall
uncertainty to $\sim 10\%$.

Overall, the above findings lead to the conclusion that the blue 
side of the HB of NGC 185 is populated by the progeny of
stars of mass slightly above $\rm 0.8~M_{\odot}$, older than
$\sim 13$ Gyr (see Fig.~\ref{ftempi}), formed
during the early evolutionary phases of the galaxy, with
current masses clustering around $\rm 0.6~M_{\odot}$. The red
side of the HB is populated the progeny of 
$\rm 0.85~M_{\odot}-0.9~M_{\odot}$ stars, with
current masses $\rm 0.65~M_{\odot}-0.7~M_{\odot}$,
with ages between 10 and 12 Gyr (Fig.~\ref{ftempi}).

\begin{figure}
\begin{minipage}{0.48\textwidth}
\resizebox{1.\hsize}{!}{\includegraphics{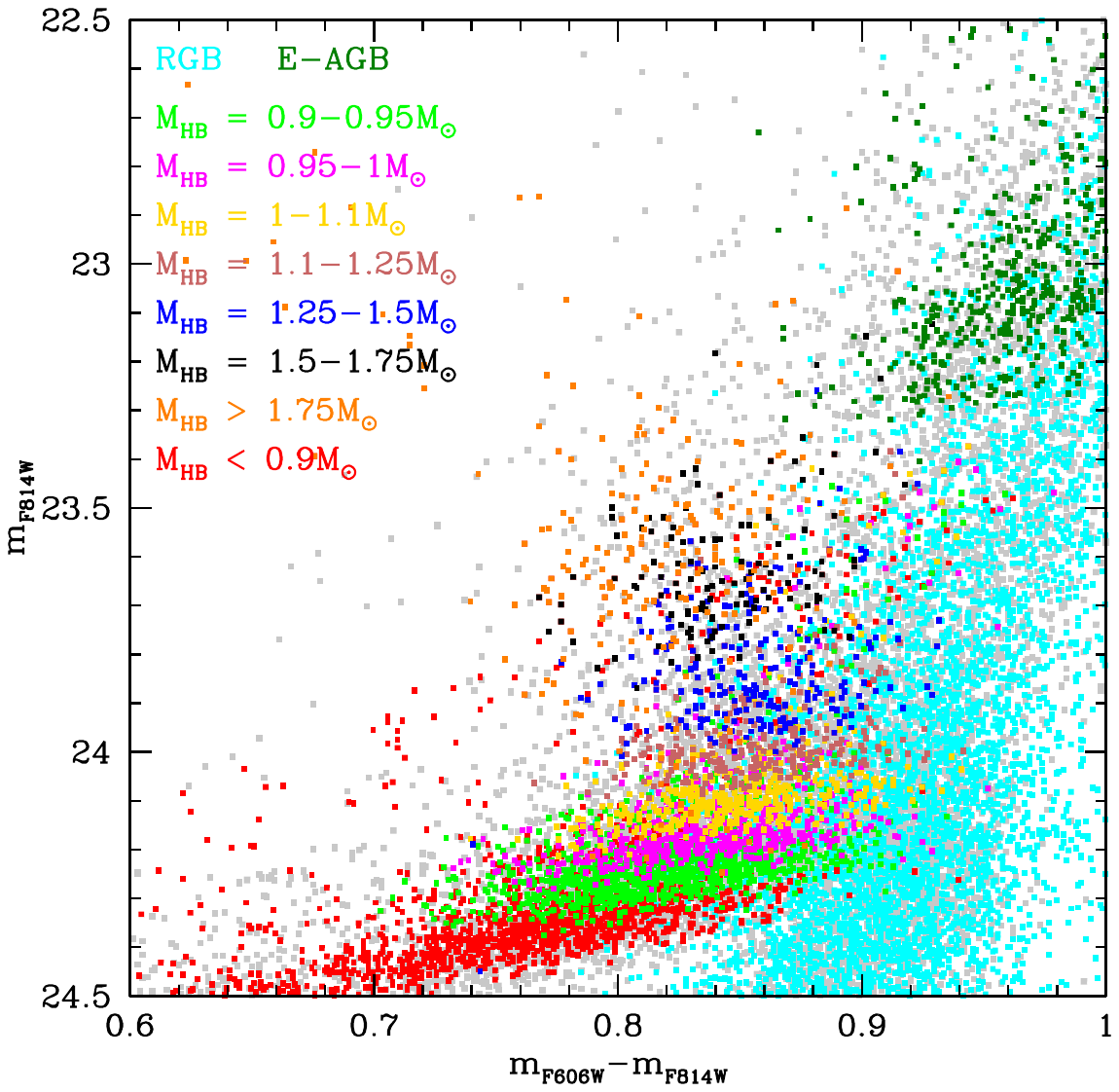}}
\end{minipage}
\vskip-60pt
\caption{Distribution of stars belonging to the galaxy NGC~185 
on the $\rm (m_{F606W}-m_{F814W}, m_{F814W})$ diagram, obtained
by the population synthesis approach described in section
\ref{sintmod}, is shown with coloured points, overlaid 
to the data, displayed in grey. Different colour coding 
refers to the mass at the start of the core helium burning
phase, as indicated in the legend.
}
\label{fclump}
\end{figure}

\subsection{RC stars witnesses of the $t<7$ Gyr epochs }
While the study of the HB of NGC 185 allows
us to reconstruct the SFH of the galaxy for ages older than $\sim 8$ Gyr,
the SFR in more recent epochs can be
assessed by analyzing the distribution of the sources populating
the RC (point (c) above).
This is made possible by the fact that the position of RC
stars on the CMD is highly
sensitive to the stellar mass, which, in turn, is related
to the formation epoch. This can be seen 
in Fig.~\ref{ftracce}, where we note that the ZAHB develops
almost vertically for $\rm M_{HB} > 0.9~M_{\odot}$ stars.

To better clarify this point we show in Fig.~\ref{fclump}
a zoom of the results reported in the right panel of
Fig.~\ref{fcmd}, focused on the RC region. The stars
of different mass are shown with different colour codes.
The $\rm m_{F814W}$ luminosity
function (LF) of stars in the 
$\rm 23~\rm{mag} < m_{F814W} < 24.5~\rm{mag}$ range
offers a notable opportunity to reconstruct the SFH
of the galaxy between $\sim 1$ and $\sim 7$ Gyr ago,
when the stars of
(current) mass in the $\rm 0.9-2~M_{\odot}$ range
formed.
Consistency between the observed and the synthetic
LF is obtained by assuming a SFR 
of $\rm \sim 5\times 10^{-4}~M_{\odot}/$yr,
8 Gyr ago, which gradually decreased, until almost
vanishing, around 800 Myr ago. The SFH
experienced during the epochs between 1 and 8 Gyr ago
can be reconstructed with an accuracy of $\sim 20\%$.
Within the present framework it is not possible 
to reconstruct with reasonable accuracy the SFH that took 
place during the last $\sim 800$ Myr, and then to test the 
claim by \citet{ren21} of some recent star formation in the
galaxy. Indeed the RGB and core helium burning phases of 
$\rm M \geq 3~M_{\odot}$ stars, which evolve on time scales 
$\rm \tau_{ev} < 800$ Myr, are so short that only a few of these
objects are expected to be observed in the CMD, thus preventing
a reliable statistical analysis.

\begin{figure}
\begin{minipage}{0.48\textwidth}
\resizebox{1.\hsize}{!}{\includegraphics{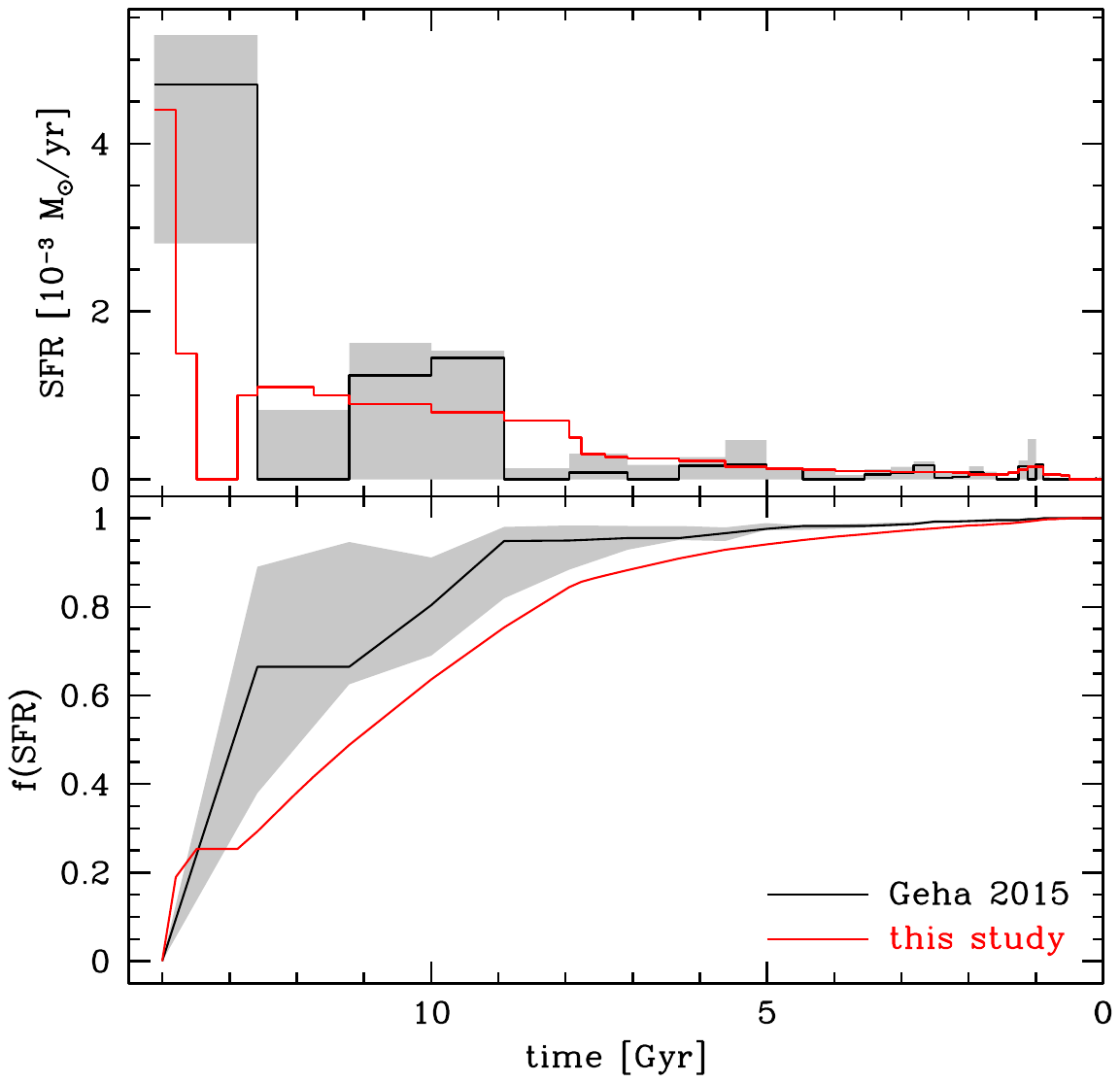}}
\end{minipage}
\vskip-60pt
\caption{The total (top panel) and the cumulative (bottom) SFR 
of NGC 185 obtained in the present work, indicated
with red lines, are compared with the results from \citet{geha15},
where grey shading indicates the combination of the systematic
and statistical uncertainties.
}
\label{fsfh}
\end{figure}

The SFH derived by the interpretation of the distribution of the
stars in the CMD is shown in Fig.~\ref{fsfh}. We note the initial
period of star formation, during which the stars populating
the blue side of the HB formed, the following break of $\sim 1$ Gyr, 
responsible for the drop in the number of HB sources in the
$0.4$ mag  $\rm < m_{F814W} < 0.6$ mag strip (see Fig.~\ref{fhb}), 
and the resumption of star formation until a few Gyr ago, during 
which the stars populating the RC formed. 

\subsection{An overall view of the luminosity function}
\begin{figure}
\begin{minipage}{0.48\textwidth}
\resizebox{1.\hsize}{!}{\includegraphics{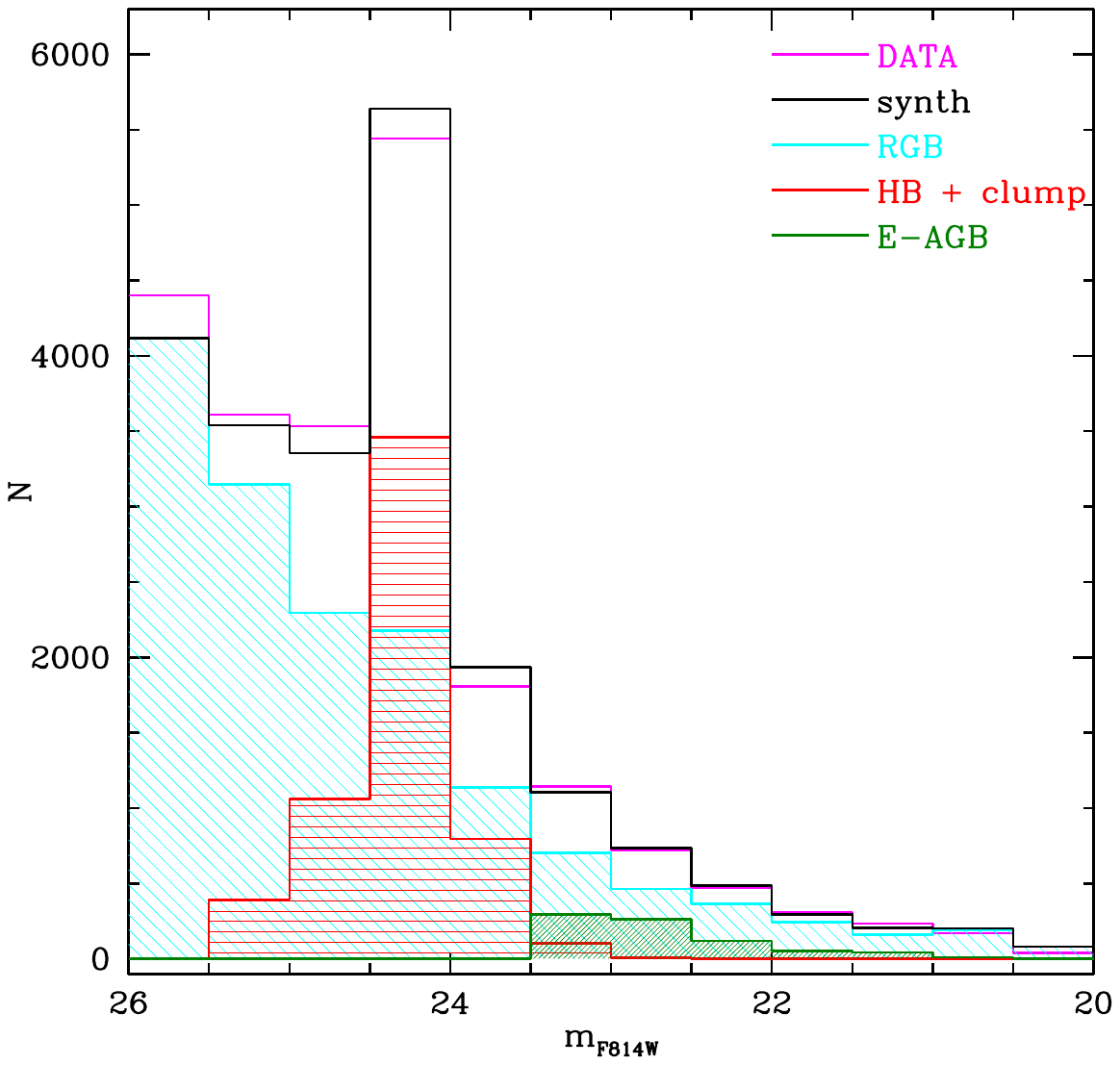}}
\end{minipage}
\vskip-60pt
\caption{Observed (magenta) $\rm m_{F814W}$ LF of
NGC~185 stars compared to the synthetic LF (black).
The separate contributions from RGB, E-AGB and core
helium burning stars are indicated with different colour
coding.
}
\label{fLF}
\end{figure}

The derivation of the SFH mostly based on core helium burning
stars described earlier in this section must be tested
against the whole dataset of NGC 185, including RGB stars,
which provide the largest contribution to the overall 
evolved population. The most straightforward way to
run this test is by comparing the synthetic $\rm m_{F814W}$
LF with that derived from stellar counts in the various
$\rm m_{F814W}$ bins. The comparison, reported in
Fig.~\ref{fLF}, shows that the agreement is extremely 
satisfactory, across the whole CMD. 

In the different $\rm m_{F814W}$ bins 
we also show the separate contributions of RGB, HB$/$RC and 
E-AGB stars to the global statistics.
Core helium burning stars are spread across the
$\rm 23 ~ mag < m_{F814W} < 25.5 ~ mag$ region, and provide the
dominant contribution in the $23.5-24$ mag $\rm m_{F814W}$ range,
where the red side of the HB joins the fainter part
of the RC, thus determining a number of stars in
excess of the RGB counterparts, whereas RGB stars dominate the 
stellar counts over the rest of the CMD. 

E-AGB stars populate a region brighter than HB and RC
stars, extending from $\rm m_{F814W} = 23.5$ mag
to $\rm m_{F814W}=21$ mag, mostly concentrated towards the 
faint side, in agreement with the discussion in section
\ref{evol}. 
It is evident the clustering of E-AGB stars in the 
$\rm 22.5 ~ mag < m_{F814W} < 23.5 ~ mag$ region of the CMD,
due to the temporary slowdown of the increase in the E-AGB luminosity 
following the extinction of the CNO burning shell 
(see Fig.~\ref{feagb}, and the relative discussion in section \ref{evol}).

\subsection{A metallicity spread?}
The results presented so far are based on stellar evolution models with
metallicity $\rm Z=2\times 10^{-3}$. We now test whether a different
chemical composition, or a metallicity spread, is compatible with the
data. The two age groups are constrained by different parts of the CMD, so
we treat them separately: the old population, which populates the HB, is
tested first (Sect.~\ref{old}); the younger population, which does not
reach the HB and is probed by the bright RGB, is tested second
(Sect.~\ref{young}).

\subsubsection{Old population: constraints from the HB}
\label{old}
\begin{figure*}
\vskip-40pt
\begin{minipage}{0.44\textwidth}
\resizebox{1.\hsize}{!}{\includegraphics{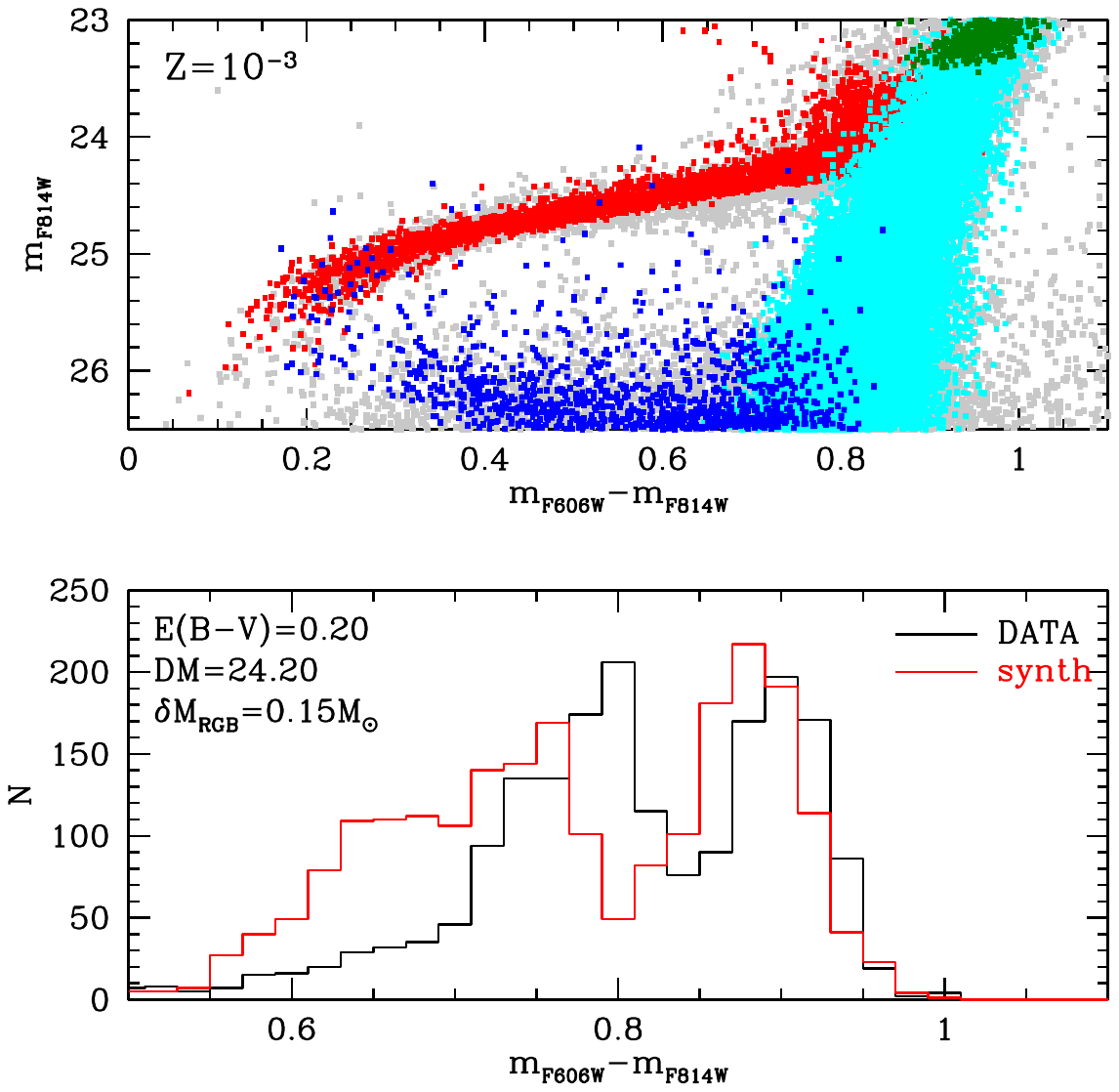}}
\end{minipage}
\begin{minipage}{0.44\textwidth}
\resizebox{1.\hsize}{!}{\includegraphics{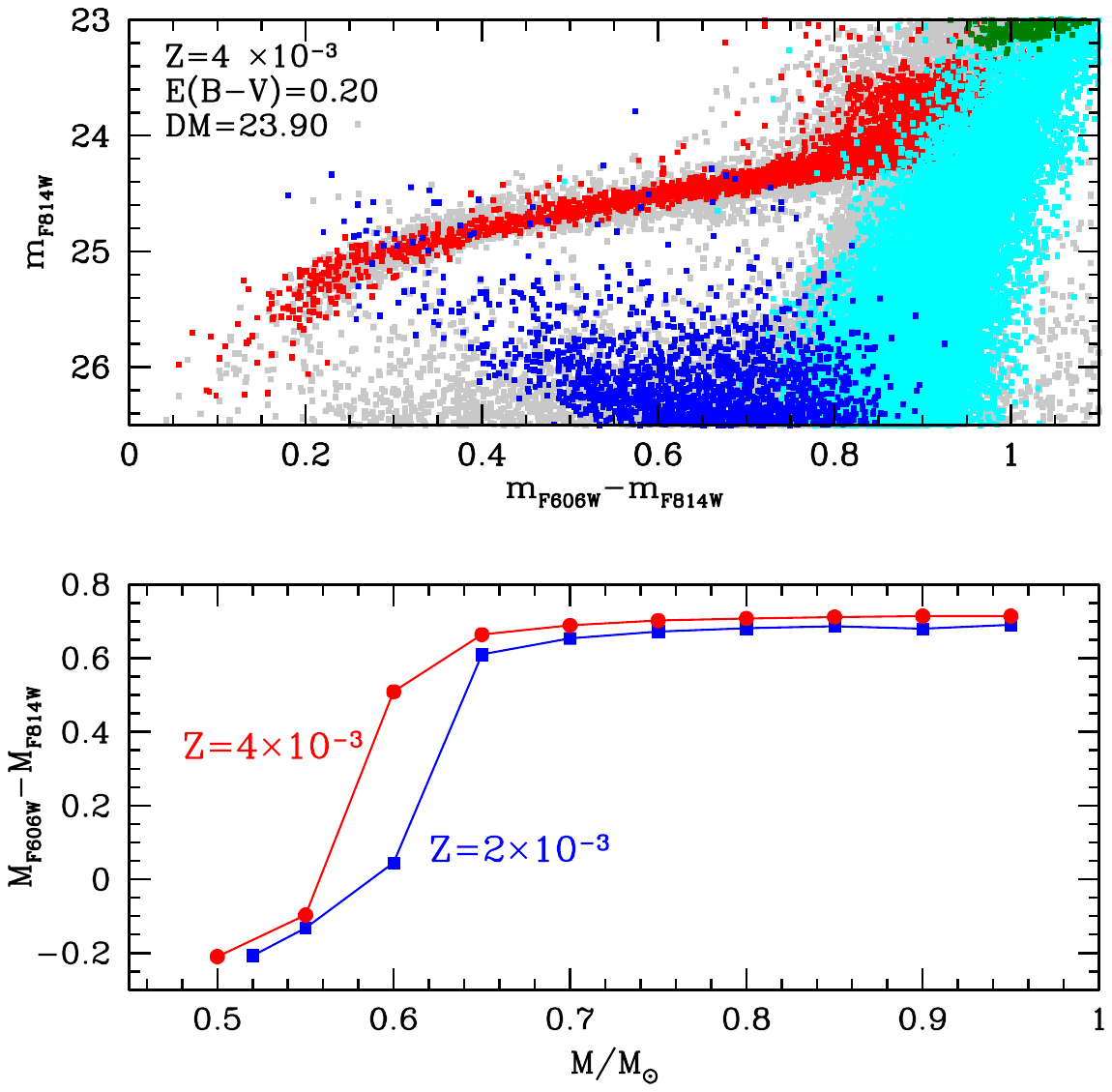}}
\end{minipage}
\vskip-60pt
\caption{Distribution of NGC 185 stars on the colour-magnitude
$\rm (m_{F606W}-m_{F814W},m_{F814W})$ diagram, compared to
results from population synthesis obtained by assuming metallicity
$\rm Z=10^{-3}$ (top, left panel) and $\rm Z=4\times 10^{-3}$ 
(top, right). The bottom, left panel shows the comparison
between the observed and the synthetic colour distribution
of HB and clump stars obtained in the $\rm Z=10^{-3}$ case,
while the colour-mass relationships of HB stars of
metallicity $\rm Z=2\times 10^{-3}$ and $\rm Z=4\times 10^{-3}$
are compared in the bottom, right panel.
} 
\label{fzeta}
\end{figure*}

We repeated the population synthesis analysis described in
Sect.~\ref{input} adopting $\rm Z=10^{-3}$ and $\rm Z=4\times 10^{-3}$,
which bracket the recommended value of $\rm Z=2\times 10^{-3}$ to within a
factor of 2. The results are shown in Fig.~\ref{fzeta}. In both cases the
distance modulus and the RGB mass loss must be readjusted to fit the blue
HB, since a lower (higher) metallicity implies a brighter (fainter) HB and
smaller (larger) turn-off masses. The reddening needs no changes
during this exploration, because the colours of the bluest HB stars
are metallicity-independent, and thus fitting the position of those object requires the same $\rm E(B-V)$.

\textit{Case $\rm Z=10^{-3}$.} The HB is brighter than in the reference
case, so a longer distance modulus, $\rm DM=24.20$ mag, is required. The
turn-off masses are smaller, so the RGB mass loss needed to reproduce the
blue HB tail is lower, $\rm \delta M_{RGB} = 0.15~M_{\odot}$. With these
parameters the blue HB is satisfactorily reproduced, at least
qualitatively (Fig.~\ref{fzeta}, top left panel). However, the synthetic
distribution fails on the red side of the HB, in the
$\rm 0.6 < m_{F606W}-m_{F814W} < 0.8$ region. This is confirmed by the
colour distribution in the bottom left panel, where the peak of the
synthetic HB is $\sim 0.05$ mag bluer than observed.

\textit{Case $\rm Z=4\times 10^{-3}$.} The HB is fainter than in the
reference case, so a shorter distance modulus, $\rm DM=23.90$ mag, is
required. The turn-off masses are higher, so a larger RGB mass loss,
$\rm \delta M_{RGB} = 0.24~M_{\odot}$, is needed to reproduce the blue HB.
The main problem here is the very small number of synthetic HB stars in
the $\rm 0.2 < m_{F606W}-m_{F814W} < 0.4$ region (top right panel). This
follows from the slope of the mass--colour relation of the
$\rm Z=4\times 10^{-3}$ ZAHB: as the bottom right panel shows, this
relation is steeper than for $\rm Z=2\times 10^{-3}$, so the range of
masses falling in that colour interval is considerably narrower.
Reproducing the observed counts would require a substantial increase in
the star formation rate between 8 and 12 Gyr ago, which in turn leads to
an unrealistic overestimate of the number of RGB stars.

\textit{Conclusion.} Neither alternative metallicity reproduces the HB.
The vast majority, if not all, of the old population of NGC 185 is
therefore composed of stars with $\rm Z=2\times 10^{-3}$.

\subsubsection{Younger population: constraints from the bright RGB}
\label{young}

\begin{figure}
\begin{minipage}{0.48\textwidth}
\resizebox{1.\hsize}{!}{\includegraphics{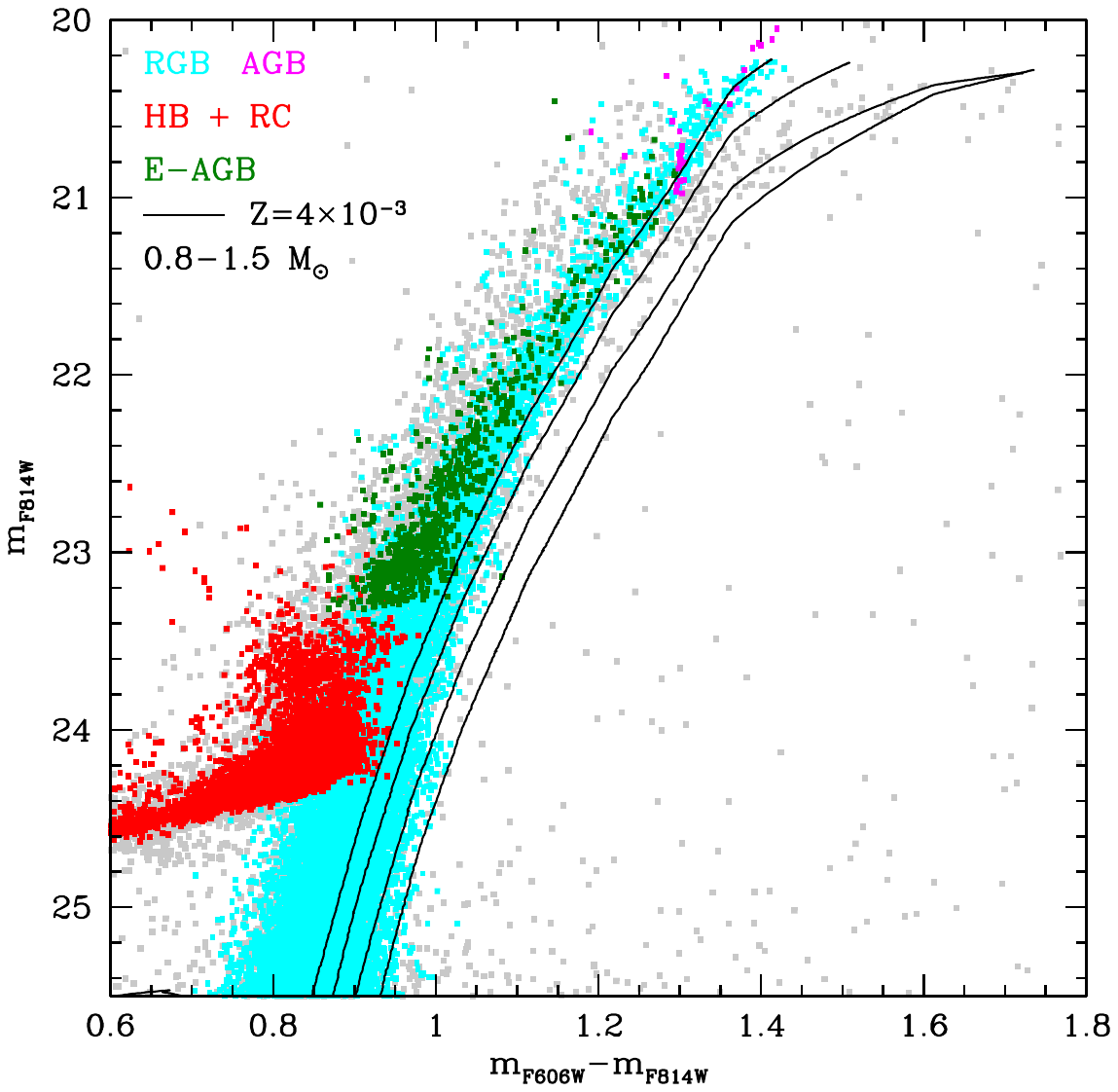}}
\end{minipage}
\vskip-60pt
\caption{The same as the right panel of Fig.~\ref{fcmd}, focused on bright
side, towards the TRGB. The black evolutionary tracks 
indicate the RGB evolution of $\rm Z=4\times 10^{-3}$
model stars of mass (from right to left) $\rm 0.8~M_{\odot}$,
$\rm 1~M_{\odot}$,$\rm 1.25~M_{\odot}$,$\rm 1.5~M_{\odot}$.
}
\label{fZ}
\end{figure}

A metallicity spread may still be present among the younger stars, which
do not populate the HB and therefore leave its morphology unaffected.
Such a metal-enriched component is suggested by the photometric analysis
of RGB stars \citep{geha10} and by spectroscopic measurements
\citep{vargas14}. To test this possibility, we examine how a
metal-enriched population would appear on the bright side of the CMD,
using the $\rm Z=4\times 10^{-3}$ evolutionary tracks considered above
(Fig.~\ref{fZ}). As expected, the tracks shift towards the blue as $\rm M$
increases, owing to the higher surface gravity.

\textit{Stars with $\rm M \leq 1~M_{\odot}$ (older than 7 Gyr).} Their
tracks lie on the red side of the observed distribution in the
$\rm m_{F814W}>21.5$ mag region, where no sources are found. A
metal-rich population older than 7 Gyr can therefore be ruled out,
consistent with the HB result above.

\textit{Stars with $\rm M > 1~M_{\odot}$ (younger than $\sim 7$ Gyr).}
Here the $\rm Z=4\times 10^{-3}$ tracks are consistent with the
observations and reproduce the position of the objects in the red, bright
part of the CMD ($\rm m_{F814W}<21.5$ mag,
$\rm (m_{F606W}-m_{F814W})>1.4$ mag). We tentatively interpret the
reddest bright sources as more metal-rich stars with masses above solar.

\textit{Limit on the metal-rich fraction.} Not all stars formed in the
last 7 Gyr can be metal-enriched. A non-negligible fraction of metal-poor
stars with masses above solar is needed to account for the bluest objects
in the brighter part of the RGB. As discussed in the first part of
Sect.~\ref{evol} and shown in the left panel of Fig.~\ref{ftracce}, the
position of these objects is reproduced only by tracks of stars with
$\rm M>1~M_{\odot}$ and metallicity $\rm Z \leq 2\times 10^{-3}$.

\subsubsection{Summary}

The old population (older than $\sim 7$ Gyr) is consistent with a single
metallicity, $\rm Z=2\times 10^{-3}$. A metal-enriched component
($\rm Z \sim 4\times 10^{-3}$) is allowed only among stars younger than
$\sim 7$ Gyr, where it can account for the reddest bright RGB sources,
but it must coexist with a substantial metal-poor fraction
($\rm Z \leq 2\times 10^{-3}$).

\section{The AGB population}
\label{agb}

Although AGB stars have little impact on the statistics of the stellar population in the ACS field 
of NGC 185, they deserve discussion because they are often used to constrain the star formation 
history (SFH) of resolved galaxies \citep{harmsen23,Lee2025}.

Our simulations predict that the ACS field contains only $\sim 70$ stars currently evolving through 
the AGB phase. These stars, shown as magenta points in the right panel of Fig.~\ref{fcmd}, populate 
the bright part of the CMD, at $\rm m_{F814W} < 21$ mag. This is expected, since they are currently 
evolving at luminosities above $\rm 10^3~L_{\odot}$, the minimum required for the onset of thermal 
pulses. Most of these AGB stars descend from progenitors of $\rm 0.8-0.9~M_{\odot}$ that formed more 
than $\sim 8$ Gyr ago, whereas a small fraction ($\sim 10$) originate from more massive progenitors 
younger than $\sim 5$ Gyr.

Observationally, fainter AGB stars in NGC 185 ACS field are largely outnumbered by RBG stars 
with similar luminosities. On the other hand, the brighter AGB stars in the field
are saturated and hence do not enter into the list of photometric stars in the ACS field 
derived in the catalogue of \cite{savino25}.

Recently, \citet{harmsen23} suggested that the ratio between the number of AGB stars within 
the region of the CMD located 0.15 mag above the TRGB, and the RGB counterparts contained in a 
box of the same height (0.6 mag) just below the TRGB, is directly correlated with T90. \citet{ventura26} 
derived an empirical relation, based on results from stellar evolution, dust formation and population 
synthesis modelling, between T90 and  $\rm N_{AGB}/N_{RGB}$. In the case of NGC 185, we find that 
the number of stars in the RGB box defined by \citet{harmsen23} is $\sim 170$, while those expected 
in the AGB box are 14, which leads to $\rm N_{AGB}/N_{RGB} \sim 0.08$. This result is in agreement with 
the observations derived from HST fields closer to the center using a combination of ACS and WFC3.
Furthermore, the application of the  $\rm T90-N_{AGB}/N_{RGB}$ relation by \citet{ventura26} gives 
$\rm T90 \sim 6.5$ Gyr, in agreement with the results shown in Fig.~\ref{fsfh}.

\begin{figure}
\begin{minipage}{0.48\textwidth}
\resizebox{1.\hsize}{!}{\includegraphics{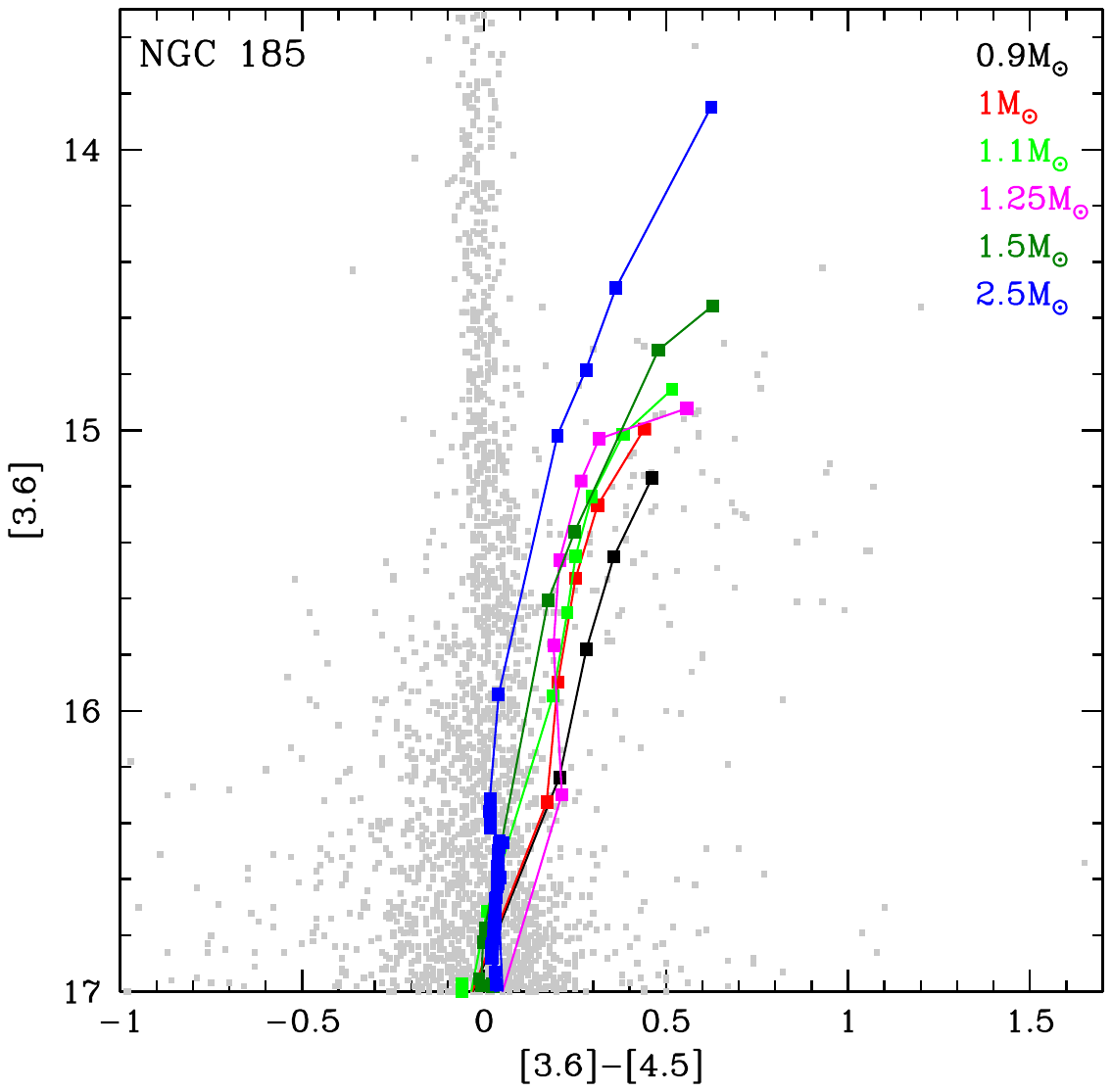}}
\end{minipage}
\vskip-60pt
\caption{The position of NGC~185 stars in the colour-magnitude
$\rm ([3.6]-[4.5], [3.6])$ diagram, based on IRAC photometry, is
indicated with grey points. Coloured points and lines indicate
the AGB evolutionary tracks of stars of different mass. Each point
along the tracks represent an evolutionary stage taken in the
middle of each inter-pulse phase.
}
\label{firac}
\end{figure}

The study of AGB stars in the galaxy is tightly connected 
to the dust production issue, as AGB stars are by far the
most efficient dust manufacturers within the evolved population.
Several dedicated studies of Local Group galaxies showed that
most of the total current dust production rate is
due to a few sources, now evolving through the final AGB 
phases \citep{flavia16, flavia18, flavia19, cla26}. 
The study of dusty stars of galaxies is preferentially addressed in
the infrared, because the presence of dust in the circumstellar
envelope makes the SED to shift from the optical to
the infrared spectral region. This is the reason why 
Spitzer data were extensively used to investigate dusty 
stars in Local Group galaxies
\citep{Boyer2011, boyer2015}, and the motivation to
study the distribution of NGC 185 stars on the
CMD obtained with Spitzer colours.

Fig.~\ref{firac} shows the distribution of NGC 185 stars on the 
$\rm ([3.6]-[4.5],[3.6])$ CMD,
obtained with the Spitzer-IRAC filters centered at $3.6~\mu$m 
and $4.5~\mu$m. As far as dust production is concerned,
the most interesting sources are those located on the
bright, red side of the diagram, in the region at
$\rm 14.5 ~mag < [3.6] < 16~mag$ and 
$\rm 0.2 ~mag < [3.6]-[4.5] < 1~mag$.
Overlapped to the data points we report the evolutionary
sequences of stars of initial mass above $\rm 0.9~M_{\odot}$,
of the same metallicity adopted in the previous sections.
The tracks of $\rm M < 0.9~M_{\odot}$ stars are not shown
in the figure, because they do not reach the C-star stage,
and thus they evolve almost vertically on the diagram,
with negligible dust production.

The results shown in Fig.~\ref{firac} indicate that the stars on the
red side of the diagram are the progeny of 
$\rm 0.9 - 1.5~M_{\odot}$ stars, formed between 1.5 and 8 
Gyr ago. These stars have become carbon stars under the
action of repeated TDU events, and have nowadays 
reached surface $\rm C/O$ ratios in the $2-5$ range.
They are currently producing dust, mostly amorphous
carbon, with rates up to a few $\rm 10^{-8}~M_{\odot}/yr$.

\section{Discussion}
\label{disc}
The analysis presented and discussed in section \ref{sfh}
demonstrated that the study of the morphology of the HB 
is the most valuable indicator of several 
properties of galaxies, opening the way to the simultaneous derivation of 
reddening $\rm E(B-V)$ and distance modulus $\rm DM$, and to the 
reconstruction of the ancient SFH, particularly in the
epochs older than $\sim 8$ Gyr. In the case of NGC 185 we
deduce that the old stellar component of the galaxy formed
during two epochs, separated by a $\sim 1$ Gyr period with 
little or no activity. The best fit of the data is obtained
with $\rm E(B-V)=0.20$ mag and $\rm DM=24.10$ mag, in substantial 
agreement with previous estimates obtained with different 
methods \citep{javadi17,savino25}. 

An important result that can be obtained by the
study of the distribution of HB stars on the CMD is
the evaluation of the average mass loss experienced
by low-mass stars while ascending the RGB. Indeed 
$\rm \delta M_{RGB}$
determines the mass with which the stars start the 
core helium burning phase, and thus affects directly
their position on the CMD, and particularly the blue
extension of the HB. The RGB mass loss has also deep
consequences on the AGB evolution: higher
values of $\rm \delta M_{RGB}$ make the stars to start
the AGB with a lower mass stored in the envelope,
so that the number of thermal pulses experienced is
smaller, and the growth of the core mass and of the
luminosity are consequently reduced.

The importance of $\rm \delta M_{RGB}$
for the interpretation of the distribution of the
evolved stellar population of galaxies on the CMD
was recently stressed by \citet{cla25}, who showed 
that the LF of the oldest AGB population of Andromeda could be reproduced
only by assuming an RGB mass loss in excess of
$\rm 0.2~M_{\odot}$. A similar result was obtained
by \citet{ventura26}, who showed that consistency 
between the expected and the derived $\rm N_{AGB}/N_{RGB}$
ratio of several galaxies with known SFH is obtained
with $\rm \delta M_{RGB}$ in the $\rm 0.2-0.25~M_{\odot}$
range. 
In the present study 
we obtained that $\rm \delta M_{RGB}=0.22~M_{\odot}$
is required to fit the blueward extension of the
HB of NGC 185. This is a welcome result, as it is
in full agreement with the analysis by \citet{ventura26},
where $\rm \delta M_{RGB}$ was derived with a 
different approach, based on the results from RGB
and AGB evolution modelling.

This result needs to be framed in the current discussion in the literature
regarding the integrated mass loss along the RGB as estimated with several 
techniques. This topic has a long history for globular clusters, dwarf galaxies 
and field stars, and many authors have attempted a thorough estimate of 
this parameter, often with conflicting results. Regarding the stellar populations 
of the galactic globular clusters the problem has been studied for a long time. 
A few examples of the results can be found in 
\cite{gratton10,origlia14,mcdonald15,tailo20,tailo21}. All these authors found 
the absolute value of the integrated mass loss to be increasing with metallicity 
and reaching, at $\rm [Fe/H]=-1$, $\rm \delta M_{RGB}\sim 0.20\, M_\odot$, and 
maximum around $\rm \delta M_{RGB}\sim 0.25\, M_\odot$, with the exact value 
depending on the source and the method. Similar estimates can be derived using 
other dwarf galaxies; examples can be found in \cite{salaris13} and \cite{savino19} 
for Sculptor and Tucana dwarf galaxies. Our results therefore is perfectly in 
line with the aforementioned sources in the literature.
We need to take into account that almost all these works compare stellar models 
with data and that each result can be affected by systematics, though, since the 
variety of models and techniques employed, the risk is lower than taking into 
account a single result.   

Several authors have recently estimated integrated mass loss using asteroseismology. 
If oscillation power can be detected in both the RGB stars of a population and their 
RC counterparts, the integrated mass loss follows from the difference between the masses 
derived for the two groups. Using data from the K2 mission \citep{howell14k2}, \citet{tailo22} 
and \citet{howell2022} independently estimated the integrated RGB mass loss in M4 and found 
compatible values, averaging $\delta M_{\rm RGB} \sim 0.20\,M_\odot$. Further work by 
\citet{howell2024,howell2025} on the clusters M80 and M9 yielded integrated mass-loss 
values in line with the trends previously reported in the literature. Our results are 
therefore also consistent with the asteroseismic estimates for metal-poor globular 
clusters observed with K2.

Integrated mass-loss values derived for less dense, less concentrated stellar populations, 
such as open clusters and field stars, appear to be in tension with those found in older, 
denser environments. Recently and independently, \citet[][and references therein]{brogaard24} 
and \citet[][and references therein]{li25} estimated the integrated mass loss of field stars 
in the Kepler \citep{koch2010} and K2 fields. Both find that the trend with metallicity inverts 
at around [Fe/H] $\approx -0.5$, with the mass loss falling to almost zero in the metal-rich 
open clusters NGC 6791 and M67. The former study finds a lower mass loss at the inversion point 
than the latter. Since it is unrealistic to assume a single value, or a single trend, across 
the whole metallicity and age range, additional factors must be at play. We therefore strongly 
advocate the development of new physical prescriptions for mass loss that can explain the trends 
in old, dense environments as well as in open clusters and field stars.

The route followed to reconstruct the SFH of NGC 185
moves from the oldest times, with the analysis of the HB,
to more recent epochs, based on the study of the RC.
Indeed the LF of RC stars is highly sensitive to the
star formation process occurred at ages younger than
$\sim 8$ Gyr, because the luminosity of the stars in
the RC is sensitive to the stellar mass: moving from the
faint to the bright side of the RC offers the possibility
of testing the SFR during younger
and younger epochs, until reaching $\sim 1$ Gyr ago.

The number counts of evolved stars in the $\rm 22 < m_{F814W} < 23.2$ 
region of the CMD is affected by the contribution
from the E-AGB population, which account for $\sim 30\%$ of
the total stellar population: 
the presence of E-AGB cannot be neglected in any statistical
analysis of the distribution of evolved stars on the
CMD, particularly in the part of the RGB from 1 to 2
magnitudes below the TRGB.

When applied to NGC~185, the methodology described above 
allowed us to derive the SFH reported in Fig.~\ref{fsfh}, 
where we note that the derived SFH is less concentrated in the
very early times when compared to the results obtained
by \citet{geha15}: this is mainly due to low star formation
rate during the epochs around 13 Gyr ago, when, on the contrary,
the star formation activity predicted by \citet{geha15} was 
at the maximum level. The present results are in much better
agreement with \citet{javadi17}, who found that the 
T90 for NGC 185 lies between 3.5 and 6 Gyr. 
Interestingly, when the value of T90 derived in the 
present study, slightly below 7 Gyr, is applied to the
$\rm T90$ vs $\rm N_{AGB}/N_{RGB}$ relationship by \citet{ventura26},
we find an $\rm N_{AGB}/N_{RGB}$ ratio fully consistent
with that derived from the observations; the same agreement
is not found when the T90 based on the results from \citet{geha15}, 
of the order of $\sim 9.5$ Gyr, is adopted.

%-----------------------------------------------------------------
\section{Conclusions}
\label{concl}

We have investigated to what extent the evolved stellar populations of 
the dwarf elliptical galaxy NGC 185 (HB, RC, RGB and AGB stars) can be 
used to reconstruct its star formation history (SFH), motivated by the 
need for techniques that reach distances of 10-15 Mpc, where the main 
sequence turn-off is beyond the reach of HST and JWST. Our approach compares 
synthetic distributions built from stellar evolution models with the HST 
colour-magnitude diagram $(m_{\rm F606W}-m_{\rm F814W},\, m_{\rm F814W})$.

The colour and magnitude distribution of HB stars proved to be an efficient 
diagnostic. It constrains the star formation older than $\sim 8$ Gyr, and 
yields a reddening $E(B-V)=0.20$ mag and a distance modulus $DM=24.10$ mag, 
both consistent with previous determinations based on independent methods.

The HB also provides the integrated mass lost by low-mass stars along the RGB, 
which we find to be $\rm \delta M_{RGB}=0.22\,M_\odot$. This quantity is the 
essential prerequisite for using evolved stars as SFH diagnostics, and has 
historically been the main source of uncertainty in HB- and RC-based analyses. 
Our value agrees with studies of globular cluster HBs and with recent results 
from the AGB luminosity function in galaxies with statistically significant 
AGB populations, such as Andromeda. With this constraint in place, the luminosity 
function of RC stars, which is highly sensitive to the mass (and hence age) 
distribution of the descendants of $\rm 1-2\,M_\odot$ stars, allowed us to reconstruct 
the SFH over the last $\sim 7$ Gyr.

The AGB population was then examined through IRAC-Spitzer photometry. A few 
AGB stars are surrounded by circumstellar dust, and comparison with stellar 
evolution and dust formation models indicates that these red objects are 
carbon stars descending from $\rm 1-1.5\,M_\odot$ progenitors, currently losing 
dust at rates not exceeding a few $\rm 10^{-8}\,M_\odot\,{\rm yr}^{-1}$.

As a further test, we derive T90 $\sim 7$ Gyr for NGC 185, while the ratio of AGB 
to RGB stars within the boxes of \citet{harmsen23} is $\rm N_{\rm AGB}/N_{\rm RGB}\sim 0.08$. 
These values follow the recently proposed T90 versus $\rm N_{\rm AGB}/N_{\rm RGB}$ relation, 
supporting the use of AGB and RGB number counts as SFH constraints for dwarf elliptical galaxies.

Taken together, our results show that, once RGB mass loss is adequately constrained,
the entire evolved stellar population can provide a time-resolved SFH for NGC 185 
without relying on the main sequence turn-off. Because this method relies on bright stars, 
it is potentially applicable to distant systems. Future work will apply it to other dwarf 
elliptical galaxies to assess its general validity, and to quantify the remaining uncertainties 
associated with the underlying stellar models.

\begin{acknowledgements}
      PV acknowledges support by the INAF-Theory-GRANT 2022 “Understanding mass loss and dust production from evolved stars”.\\
      CV acknowledges support by the INAF-fellowship 2026 “Web services and imaging data reduction for LBT” within the INAF project “Supporto Arizona \& Italia”.
\end{acknowledgements}

% WARNING
%-------------------------------------------------------------------
% Please note that we have included the references to the file aa.dem in
% order to compile it, but we ask you to:
%
% - use BibTeX with the regular commands:
%   \bibliographystyle{aa} % style aa.bst
%   \bibliography{Yourfile} % your references Yourfile.bib
%
% - join the .bib files when you upload your source files
%-------------------------------------------------------------------

\end{document}